\documentclass[aps,prl,twocolumn,superscriptaddress,showpacs,preprintnumbers,amsmath,amssymb]{revtex4-2}

\usepackage{orcidlink}
\usepackage{graphicx} % Include figure files
\usepackage{dcolumn}  % Align table columns on decimal point
\usepackage{lineno}
\usepackage{color}
\usepackage[normalem]{ulem}

\def\dbar{\overline{D}{}^{\,0}}

\def\Dkskspp{D^0\ra K^0_S\,K^0_S\,\pi^+\pi^-}
\def\Dksksks{D^0\ra K^0_S\,K^0_S\,K^0_S}

\def\kskspp{K^0_S\,K^0_S\,\pi^+\pi^-}
\def\Dkspp{D^0\ra K^0_S\,\pi^+\pi^-}
\def\Dkk{D^0\ra K^+\,K^-}
\def\Dpp{D^0\ra\pi^+\pi^-}
\def\Dkp{D^0\ra K^-\pi^+}
\def\Dkpo{D^0\ra K^0_S\pi^0}
\def\DstarDpi{D^{*+}\ra D^0\pi^+_s}
\def\KSpp{K^0_S\ra\pi^+\pi^-}

\def\cp{\textit{CP}}
\def\cpv{\textit{CPV}}
\def\Acp{A^{}_{\it CP}}
\def\acp{a_{\it CP}^{T}}
\def\ra{\!\rightarrow\!}
\def\Acpdet{A^{\rm{det}}_{\it CP}}
\def\Aslpi{A_{\epsilon}^{\pi_{s}}}
\def\Afb{A_{\rm FB}}
\def\Acpcor{A_{\it CP}^{\rm{cor}}}

\def\pt {p^{}_{\rm T}}
\def\costh {\cos\theta^{}_{\pi^{}_s}}
\def\cosths {\cos\theta^*}

\def\simge{\mathrel{%
   \rlap{\raise 0.511ex \hbox{$>$}}{\lower 0.511ex \hbox{$\sim$}}}}
\def\simle{\mathrel{
   \rlap{\raise 0.511ex \hbox{$<$}}{\lower 0.511ex \hbox{$\sim$}}}}

\graphicspath{{ps}}

\begin{document}

\vspace*{-3\baselineskip}
\resizebox{!}{3cm}{\includegraphics{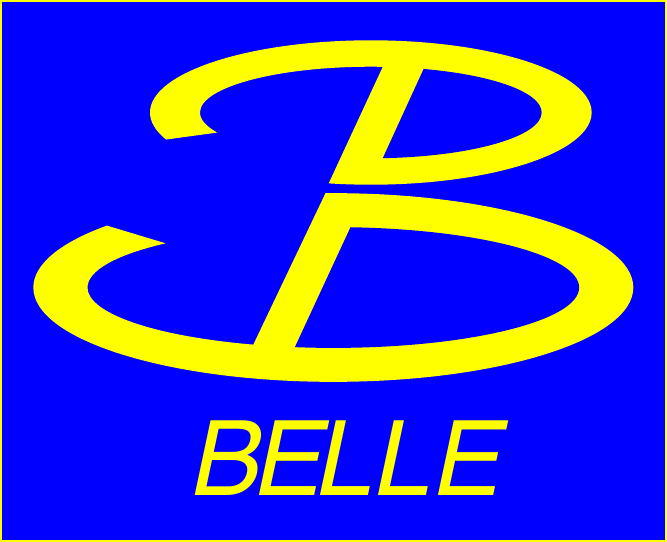}}

\vspace*{\baselineskip}
\begin{flushright}
\end{flushright}

\preprint{Belle Preprint 2022-16}
\preprint{KEK Preprint 2022-19}

\title{ %\quad\\[1.0cm] 
{\boldmath Measurement of the branching fraction and search for
\cp\ violation in $\Dkskspp$ decays at Belle}
}

%%% Paper:    D0 to KS KS pi+ pi- CPV
%%% Journal:  Physical Review D Letter
%%% Contacts: A. Sangal (amansangal9@gmail.com)
%%%           A. Schwartz (schwaraz@ucmail.uc.edu)
%%% Non-responding authors or those who said NO are commented out.
%%% ====================================================================
%%% Click the RELOAD button on your web browser to see the updated file.
%%% ====================================================================
%%% Use \input{author-orcid} to insert this material into your latex file.
%%%%% Force institutions to appear in alphabetical order when typeset.
\noaffiliation
% \author{I.~Adachi\,\orcidlink{0000-0003-2287-0173}} % 2590
% \author{K.~Adamczyk\,\orcidlink{0000-0001-6208-0876}} % 2239
% \author{J.~K.~Ahn\,\orcidlink{0000-0002-5795-2243}} % 7423
  \author{A.~Sangal\,\orcidlink{0000-0001-5853-349X}} % 2384
  \author{A.~J.~Schwartz\,\orcidlink{0000-0002-7310-1983}} % 2162
  \author{H.~Aihara\,\orcidlink{0000-0002-1907-5964}} % 2223
  \author{S.~Al~Said\,\orcidlink{0000-0002-4895-3869}} % 6823
  \author{D.~M.~Asner\,\orcidlink{0000-0002-1586-5790}} % 4684
  \author{H.~Atmacan\,\orcidlink{0000-0003-2435-501X}} % 2538
  \author{V.~Aulchenko\,\orcidlink{0000-0002-5394-4406}} % 8183
  \author{T.~Aushev\,\orcidlink{0000-0002-6347-7055}} % 3747
  \author{R.~Ayad\,\orcidlink{0000-0003-3466-9290}} % 3766
% \author{T.~Aziz\,\orcidlink{-}} % 3523
  \author{V.~Babu\,\orcidlink{0000-0003-0419-6912}} % 5623
% ...
% 4983
\collaboration{The Belle Collaboration}

\begin{abstract}
We measure the branching fraction for the Cabibbo-suppressed decay 
$D^0 \rightarrow K^0_S\,K^0_S\,\pi^+\pi^-$ and search for $CP$ violation 
via a measurement of the $CP$ asymmetry $A^{}_{CP}$ as well as the $T$-odd 
triple-product asymmetry~$a_{CP}^{T}$.  We use 922~fb$^{-1}$ of data recorded 
by the Belle experiment, which ran at the KEKB asymmetric-energy $e^+ e^-$ 
collider. The branching fraction is measured relative to the Cabibbo-favored 
normalization channel $D^0 \rightarrow K^0_S\,\pi^+\pi^-$; the result is 
$\mathcal{B}(D^0 \rightarrow K^0_S\,K^0_S\,\pi^+\pi^-) =  
[4.79 \pm 0.08\,({\rm stat}) \, \pm 0.10\,({\rm syst}) \pm 0.31\,({\rm norm})]\times 10^{-4}$, 
where the first uncertainty is statistical, the second is systematic, and the third is from 
uncertainty in the normalization channel. We also measure 
$A^{}_{CP} = [-2.51\,\pm 1.44\,({\rm stat})\,^{+0.11}_{-0.10}\,({\rm syst})]\%$, 
and $a_{CP}^{T} = [-1.95\,\pm 1.42\,({\rm stat})\,^{+0.14}_{-0.12}\,({\rm syst})]\%$. 
These results show no evidence of $CP$~violation.
\end{abstract}

\maketitle

%%%% >>>> keep the final version single-spaced \tighten

An outstanding puzzle in particle physics is the absence of antimatter observed 
in the Universe~\cite{Canetti:2012zc,Farrar:1993hn}. 
It is often posited that equal amounts of matter and antimatter
existed in the early Universe~\cite{Allahverdi_2021}.
For such an initial state to evolve into our current Universe 
requires violation of \cp\ (charge-conjugation and parity) symmetry~\cite{Sakharov:1967dj}.
Such \cp\ violation (\cpv) is incorporated naturally into the Standard Model~(SM)
via the Kobayashi-Maskawa mechanism~\cite{Kobayashi:1973fv}. However, the amount 
of \cpv\ measured to date is insufficient to account for the observed imbalance 
between matter and antimatter \cite{Farrar:1993hn,Huet:1994jb}. 
Thus, it is important to search for new sources of \cpv. 

In this paper, we search for \cpv\ in the singly Cabibbo-suppressed (SCS) 
decay $\Dkskspp$~\cite{charge-conjugates}.
SCS decays are expected to be especially sensitive to physics 
beyond the SM, as their amplitudes receive contributions 
from QCD ``penguin'' operators and also chromomagnetic 
dipole operators~\cite{Grossman:2006jg}. 
The SCS decays $\Dkk$ and $\Dpp$~\cite{LHCb:2019hro}
are the only decay modes in which \cpv\ has been 
observed in the charm sector. 
The \cp\ asymmetry measured,
\begin{eqnarray}
\Acp & \equiv & \frac{\Gamma(D^0\ra f) - \Gamma(\dbar\ra \bar{f})}
            {\Gamma(D^0\ra f) + \Gamma(\dbar\ra \bar{f})}
\end{eqnarray}
where $f$ and $\bar{f}$ are $\cp$-conjugate final states,
is small, at the level of~0.1\%.

We also perform a high-statistics measurement 
of the branching fraction.
Several measurements of the branching fraction 
exist~\cite{Albrecht:1994,FOCUS:2004met,BESIII:2020rxv}.
The most precise result was obtained 
by the BES\,III Collaboration, which found
$\mathcal{B}(\Dkskspp) =(5.3\,\pm 0.9\,\pm 0.3)\times 10^{-4}$~\cite{BESIII:2020rxv}.
Our measurement presented here uses an event sample 
almost 
two orders of magnitude larger than that of BES\,III.

We search for \cpv\ in $\Dkskspp$ decays in two 
complementary ways.
We first measure the asymmetry $\Acp$; a nonzero 
value results from interference between contributing
decay amplitudes. The \cp-violating interference term 
is proportional to $\cos(\phi + \delta)$ for $D^0$ decays, 
where $\phi$ and $\delta$ are the weak and strong
phase differences, respectively, between the amplitudes.
For $\dbar$ decays, the interference term is proportional 
to $\cos(-\phi+\delta)$. Thus, to observe a difference 
between $D^0$ and $\dbar$ decays (i.e., $\Acp\neq 0$), 
$\delta$ must be nonzero.

To avoid the need for $\delta\neq 0$, we also search for 
\cpv\ by measuring the asymmetry in the triple-product 
$C^{}_{T} = \vec{p}^{}_{K^0_S} \cdot (\vec{p}^{}_{\pi^+} \times \vec{p}^{}_{\pi^-})$,
where $\vec{p}^{}_{K^0_S}$, $\vec{p}^{}_{\pi^+}$, and $\vec{p}^{}_{\pi^-}$ are
the three-momenta of the $K^0_S$, $\pi^+$, and $\pi^-$ 
daughters, defined in the $D^0$ rest frame. We use the 
$K^0_S$ with the higher momentum for this calculation.
The asymmetry is defined as
\begin{eqnarray}
A^{}_T & \equiv &  \frac{N(C^{}_T\!>\!0) - N(C^{}_T\!<\!0)}
{N(C^{}_T\!>\!0) + N(C^{}_T\!<\!0)}\,,
\end{eqnarray}
where $N(C^{}_T\!>\!0)$ and $N(C^{}_T\!<\!0)$ correspond to the yields of $\Dkskspp$ 
decays having $C^{}_T\!>\!0$ and $C^{}_T\!<\!0$, respectively.
The observable $A^{}_T$ is proportional to 
$\sin(\phi + \delta)$~\cite{Durieux:2015zwa,PhysRevD.39.3339,Bensalem:2000hq}.
For $\dbar$ decays, we define the \cp-conjugate quantity
\begin{eqnarray}
\bar{A}^{}_T & \equiv &  \frac{\overline{N}(-\overline{C}^{}_T>0) - \overline{N}(-\overline{C}^{}_T<0)}
{\overline{N}(-\overline{C}^{}_T>0) + \overline{N}(-\overline{C}^{}_T<0)}\,,
\label{eqn:atbar}
\end{eqnarray}
which is proportional to $\sin(-\phi + \delta)$. 
Thus, the difference 
\begin{eqnarray}
\acp & \equiv & \frac{A^{}_T - \bar{A}^{}_T}{2} 
\end{eqnarray}
is proportional to $\sin\phi\cos\delta$, and, unlike $\Acp$,
$\delta=0$ results in the largest \cp\ asymmetry. 
The minus sign in front of $\overline{C}^{}_T$
in Eq.~(\ref{eqn:atbar}) corresponds to the parity 
transformation, which is needed for $\bar{A}^{}_T$ to be the \cp-conjugate 
of $A^{}_T$. Finally, we note that $\acp$ is advantageous to measure experimentally,
as any production asymmetry between $D^0$ and $\dbar$ or difference in 
reconstruction efficiencies cancels out.

We measure the branching fraction, $\Acp$, and $\acp$ using 
data collected by the Belle experiment running at the KEKB 
asymmetric-energy $e^{+} e^{-}$ collider~\cite{kekb}. 
The data used in this analysis
were collected at $e^{+} e^{-}$ center-of-mass (CM) energies 
corresponding to the $\Upsilon(4S)$ and $\Upsilon(5S)$ resonances,
and 60~MeV below the $\Upsilon(4S)$ resonance. The total 
integrated luminosity is 922~fb$^{-1}$.
% corresponding to an integrated luminosity of 711 fb$^{-1}$, 121 fb$^{-1}$ and 89 fb$^{-1}$ respectively. \\

The Belle detector~\cite{Abashian:2000cg} is a large-solid-angle magnetic
spectrometer consisting of a silicon vertex detector (SVD),
a 50-layer central drift chamber (CDC), an array of
aerogel threshold Cherenkov counters (ACC),  % <- \v{C}erenkov 2007.08
a barrel-like arrangement of time-of-flight
scintillation counters (TOF), and an electromagnetic calorimeter
comprising CsI(Tl) crystals. All these subdetectors are
located inside a superconducting solenoid coil that provides 
a 1.5~T magnetic field.  An iron flux-return located outside the 
coil is instrumented to detect $K_L^0$ mesons and to identify
muons. Two inner detector configurations were used: 
a 2.0-cm-radius beam-pipe and a three-layer SVD were 
used for the first 140~fb$^{-1}$ of data, and a 1.5-cm-radius 
beam-pipe, a four-layer SVD, and a small-inner-cell drift chamber 
were used for the remaining data~\cite{Natkaniec:2006rv}.

We use Monte Carlo (MC) simulated events to optimize event selection 
criteria, calculate reconstruction efficiencies, and study sources of background.
The MC samples are generated using the {\sc EvtGen} software package~\cite{Lange:2001uf}, 
and the detector response is simulated using {\sc Geant3}~\cite{Brun:1987ma}. 
Final-state radiation is included in the simulation via the {\sc Photos} 
package~\cite{Barberio:1993qi}. To avoid introducing bias 
in our analysis, we analyze the data in a ``blind" manner, 
i.e, we finalize all selection criteria before viewing 
signal candidate events.

We identify the flavor of the $D^0$ or $\dbar$ decay by reconstructing 
the decay chain $\DstarDpi, \,\Dkskspp$; the charge of the $\pi^\pm_s$ 
(which has low momentum and is referred to as the ``slow'' pion) determines 
the flavor of the $D^0$ or $\dbar$.
The $D^0$ and $D^{*+}$ decays are reconstructed by first selecting 
charged tracks that originate from near the $e^{+}e^{-}$ interaction
point (IP). We require that the impact parameter $\delta z$ of a track 
along the $z$ direction (antiparallel to the $e^+$ beam) satisfies 
$|\delta z| < 5.0$~cm, and that the impact parameter transverse 
to the $z$ axis satisfies $\delta r < 2.0$~cm.

To identify pion tracks, we use light yield information from the ACC, 
timing information from the TOF, and specific ionization ($dE/dx$) 
information from the CDC. This information is combined into likelihoods 
$\mathcal{L}_{K}$ and $\mathcal{L}_{\pi}$ for a track to be 
a $K^+$ or $\pi^+$, respectively. To identify 
$\pi^\pm$ tracks from $\Dkskspp$, we require 
$\mathcal{L}_{\pi}/(\mathcal{L}_{\pi}+\mathcal{L}_{K}) >0.60$.
This requirement is more than 96\% efficient and has a $K^+$ 
misidentification rate of~6\%.

We reconstruct $K^{0}_S \ra \pi^{+} \pi^{-}$ decays using 
a neural network (NN)~\cite{Dash:2017heu}. The NN utilizes 
13 input variables: 
the $K^{0}_S$ momentum in the laboratory frame;
the separation along the $z$ axis between the two $\pi^\pm$ tracks;
the impact parameter with respect to the IP transverse to the $z$ axis of 
%the distance-of-closest-approach in the $x$-$y$ plane between the IP and 
the $\pi^\pm$ tracks; the $K^{0}_S$ flight length in the $x$-$y$ plane;
the angle between the $K_S^{0}$ momentum and the vector 
joining the IP to the $K_S^{0}$ decay vertex; in the 
$K_S^{0}$ rest frame, the angle between the $\pi^+$ 
momentum and the laboratory-frame 
boost direction; and, for each $\pi^\pm$ track, 
the number of CDC hits in both stereo and axial views,
and the presence or absence of SVD hits. 
The invariant mass of the two pions is required to satisfy 
$|M(\pi^{+}\pi^{-}) - m^{}_{K_S^{0}}|<0.010$~GeV/$c^2$, where $m^{}_{K^0_S}$ 
is the $K^0_S$ mass~\cite{ParticleDataGroup:2022pth}. This range 
corresponds to three standard deviations in the mass resolution.

After identifying $\pi^{\pm}$ and $K_S^{0}$ candidates, 
we reconstruct $D^{0}$ candidates by requiring that the 
%$K^0_S\,K^0_S\pi^+\pi^-$ 
four-body invariant mass $M(K^0_S\,K^0_S\pi^+\pi^-)\equiv M$ 
satisfy $1.810~{\rm GeV}/c^2 < M < 1.920~{\rm GeV}/c^2$. 
We remove $\Dksksks$ decays, which have the same 
final-state particles,
by requiring $|M(\pi^{+}\pi^{-}) - m^{}_{K^0_S}|>0.010$~GeV/$c^2$.
This criterion removes 96\% of these decays.
To improve the mass resolution, we apply mass-constrained 
vertex fits for the $K_S^{0}$ candidates. These fits require 
that the $\pi^\pm$ tracks originate from a common point, 
and that $M(\pi^+\pi^-) = m^{}_{K^0_S}$~\cite{ParticleDataGroup:2022pth}. 
We perform a vertex fit for the $D^{0}$ candidate using the
$\pi^\pm$ tracks and the momenta of the $K_S^0$ candidates;
the resulting fit quality ($\chi^{2}$) must
satisfy a loose requirement to ensure that the tracks 
and $K^0_S$ candidates are consistent with originating 
from a common decay vertex.

We reconstruct $\DstarDpi$ decays by combining $D^0$ candidates 
with $\pi^+_s$ candidates. We require that the mass difference 
$M(K^0_S\,K^0_S\pi^+\pi^-\pi^+_s) - M\equiv \Delta M$ be less than 
0.15~GeV/$c^2$. We also require that the momentum
of the $D^{*+}$ candidate in the CM frame be greater than 2.5~GeV/$c$; 
this reduces combinatorial background and also removes $D^{*+}$ 
candidates originating from $B$ decays, which can potentially
contribute their own
\cpv~\cite{BaBar:2008xnt,BaBar:2012cyz,Belle:2012xkw,Belle:2012mef,LHCb:2019ouq}.
We perform a $D^{*+}$ vertex fit, constraining 
the $D^{0}$ and $\pi^+_s$ to originate from the IP.
%we require that the $\pi^{\pm}$ from $D^{0}$ and $\pi_{\rm{slow}}$ to have at least 
%1 and 2 hits respectively in the z and $r-\phi$ measuring layers of SVD detector.  
We subsequently require $\sum(\chi^{2}/\rm{ndf}) <100$, where 
the sum runs over the two mass-constrained $K^0_S$ vertex fits, 
the $D^0$ vertex fit, and the IP-constrained $D^{*+}$ vertex fit,
and ``ndf'' is the number of degrees of freedom in each fit.

The $D^{*+}$ momentum 
%$p^{*}_{D^{*}}$ 
and $\sum(\chi^{2}/\rm{ndf})$ requirements are chosen by maximizing 
a figure-of-merit (FOM). This FOM is taken to be the ratio
$N^{}_S/\sqrt{N^{}_S+N^{}_B}$, where $N^{}_S$ and $N^{}_B$ are the numbers 
of signal and background events, respectively, expected in the signal region 
$1.845\,{\rm GeV}/c^2 < M <1.885\,{\rm GeV}/c^2$ and 
$0.144\,{\rm GeV}/c^2 < \Delta M <0.147\,{\rm GeV}/c^2$. 
The signal yield $N^{}_S$ is obtained from MC simulation 
using the PDG value~\cite{ParticleDataGroup:2022pth} for the branching fraction, 
while the background yield $N^{}_B$ is obtained by appropriately scaling 
the number of events observed in the data sideband 
$\Delta M \in (0.140,0.143) \cup (0.148,0.150)\,{\rm GeV}/c^2$.

After applying all selection criteria, 27\% of events have
multiple $\DstarDpi,\,\Dkskspp$ signal candidates. For these
events, we retain a single candidate by choosing that with 
the lowest value of $\sum(\chi^{2}/\rm{ndf})$. 
According to MC simulation, this criterion correctly 
identifies the true signal decay 81\% of the time,
without introducing any bias.

We determine the signal yield via a two-dimensional unbinned extended 
maximum-likelihood  fit to the variables $M$ and $\Delta M$. The fitted 
ranges are $1.810~{\rm GeV}/c^2 < M < 1.920~{\rm GeV}/c^2$ and 
$0.140~{\rm GeV}/c^2 < \Delta M < 0.150~{\rm GeV}/c^2$.
Separate probability density functions (PDFs) are used 
for the following categories of events: 
{\rm (a)}~correctly reconstructed signal events;
{\rm (b)}~misreconstructed signal events, i.e., one or more daughter tracks are missing;
{\rm (c)}~``slow pion background,'' i.e., a true $\Dkskspp$ decay is combined with an 
extraneous $\pi^+_s$ track;
{\rm (d)}~``broken charm background,'' i.e., a true $\DstarDpi$ decay is reconstructed, 
but the (nonsignal) $D^0$ decay is misreconstructed, faking a $\Dkskspp$ decay; 
{\rm (e)}~purely combinatorial background, i.e., no true $D^{*+}$ or $D^0$ decay; and
{\rm (f)}~$\Dksksks$ decays that survive the $M(\pi^+\pi^-)$ veto.

All PDFs are taken to factorize as $P(M)\times P(\Delta M)$. 
We have checked for possible correlations between $M$ and $\Delta M$ 
for all the signal and background components and found them to be 
negligible. For correctly reconstructed signal decays, the PDF for $M$ 
is the sum of three asymmetric Gaussians with a common mean. The PDF 
for $\Delta M$ is the sum of two asymmetric Gaussians and a 
Student's~t function~\cite{student-t}, all with a common mean.
Both common means are floated, as are the widths of the asymmetric 
Gaussian with the largest fraction used for $M$, and the $\sigma,\,r$ 
parameters of the Student's~t function used for $\Delta M$. All other 
parameters are fixed to MC values. For misreconstructed signal decays, 
a second-order Chebychev polynomial is used for $M$, and a fourth-order 
Chebychev polynomial is used for $\Delta M$. These shape parameters are 
fixed to MC values. The yield is taken to be a fixed fraction of the total 
signal yield ($14\pm 1$\%), which is also obtained from MC simulation.

\begin{figure}[ht]
\centering
\includegraphics[width=0.51\textwidth]{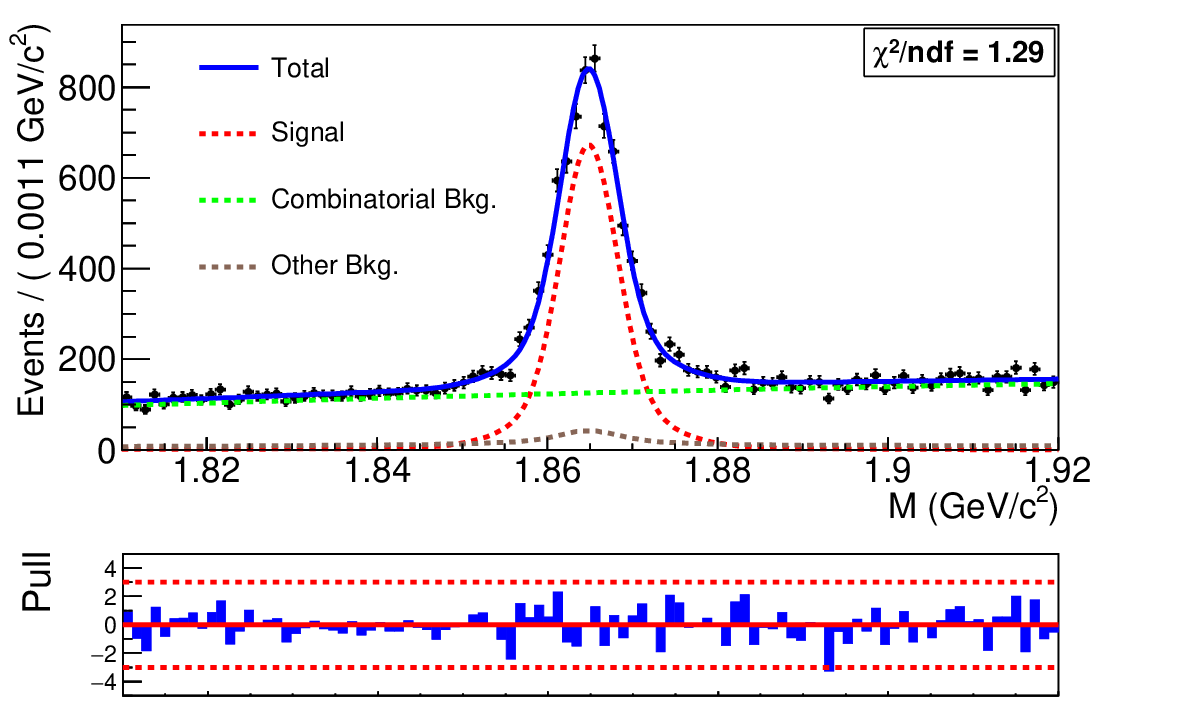}
\vspace{0.5cm}
\includegraphics[width=0.51\textwidth]{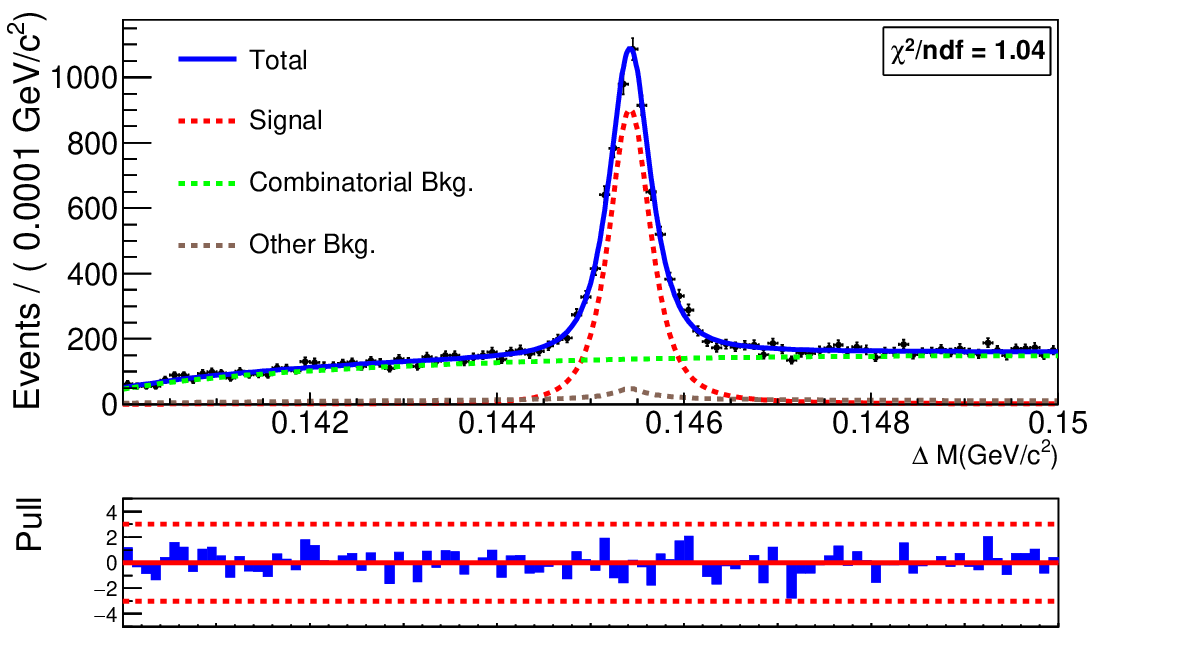}
\caption{Projections of the fit for $\Dkskspp$ on $M$ (upper) 
and $\Delta M$ (lower).
The brown dashed curve consists of
slow pion, broken charm, and $\Dksksks$ backgrounds.
The corresponding pull distributions 
[$= ({\rm data} - {\rm fit\ result})/({\rm data\ uncertainty})$] 
are shown below each projection. The dashed red lines correspond to $\pm 3\sigma$ values. 
}
\label{fig:BF_signal}
\end{figure}

For slow pion background, we use the same PDF for $M$ as 
used for correctly reconstructed signal decays. For 
$\Delta M$, we use a threshold function 
$Q^{0.5} + \alpha\cdot Q^{1.5}$, where 
$Q = \Delta M - m^{}_{\pi^+}$ and $\alpha$ is a parameter.
For broken charm background, we use the sum 
of two Gaussians with a common mean for $M$, 
and a Student's~t function for $\Delta M$. 
For combinatorial background, we use a second-order 
Chebychev polynomial for $M$, and, for $\Delta M$, 
a threshold function with the same functional form 
as used for slow pion background. For $\Dksksks$ decays, 
we use a single Gaussian for $M$ and a Student's~t function 
for $\Delta M$. The broken charm and $\Dksksks$ backgrounds 
are small; thus, their yields and shape parameters are taken 
from MC simulation. For slow pion background, the $\Delta M$ 
shape parameters are taken from MC simulation. 
All other shape parameters
(six for the means and widths of the signal PDF, 
and three for the combinatorial background) are floated. 
The fit yields $6095\pm 98$ signal events. Projections of 
the fit are shown in Fig.~\ref{fig:BF_signal}.

We normalize the sensitivity of our search by counting the number of $\Dkspp$ decays 
observed in the same dataset. The branching fraction for $\Dkskspp$ is calculated as
\begin{eqnarray}
\mathcal{B}(\Dkskspp) & = &  \nonumber \\
 & & \hskip-1.6in 
\left(\frac{N_{K^0_S\,K^0_S\pi^+\pi^-}}{N_{K^0_S\pi^+\pi^-}}\right)
\left(\frac{\varepsilon^{}_{K^0_S\pi^+\pi^-}}{\varepsilon^{}_{K^0_S\,K^0_S\pi^+\pi^-}}\right)
\times \frac{\mathcal{B}(\Dkspp)}{\mathcal{B}(\KSpp)}\,, \nonumber \\
\label{eqn:br}
\end{eqnarray}
where $N$ is the fitted yield for $\Dkskspp$ or $\Dkspp$  decays;  
$\varepsilon$ is the corresponding reconstruction efficiency, given that $K^0_S\ra\pi^+\pi^-$; 
and $\mathcal{B}(\KSpp)$ and $\mathcal{B}(\Dkspp)$ are the world average branching fractions 
for $\KSpp$ and $\Dkspp$~\cite{ParticleDataGroup:2022pth}.
The selection criteria for $\Dkspp$ are the same as those used for $\Dkskspp$, 
except that only one $K^0_S$ is required.

\begin{figure}[ht]

\centering

\includegraphics[width=0.51\textwidth]{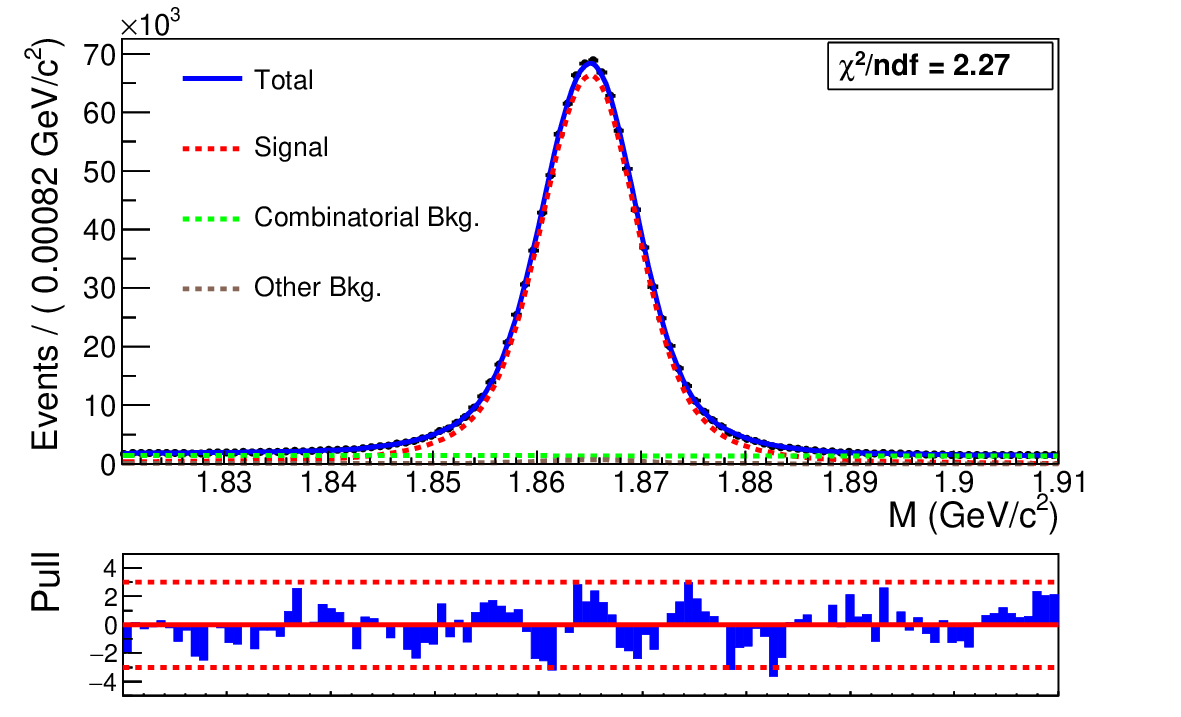}

\vspace{0.5cm}
\includegraphics[width=0.51\textwidth]{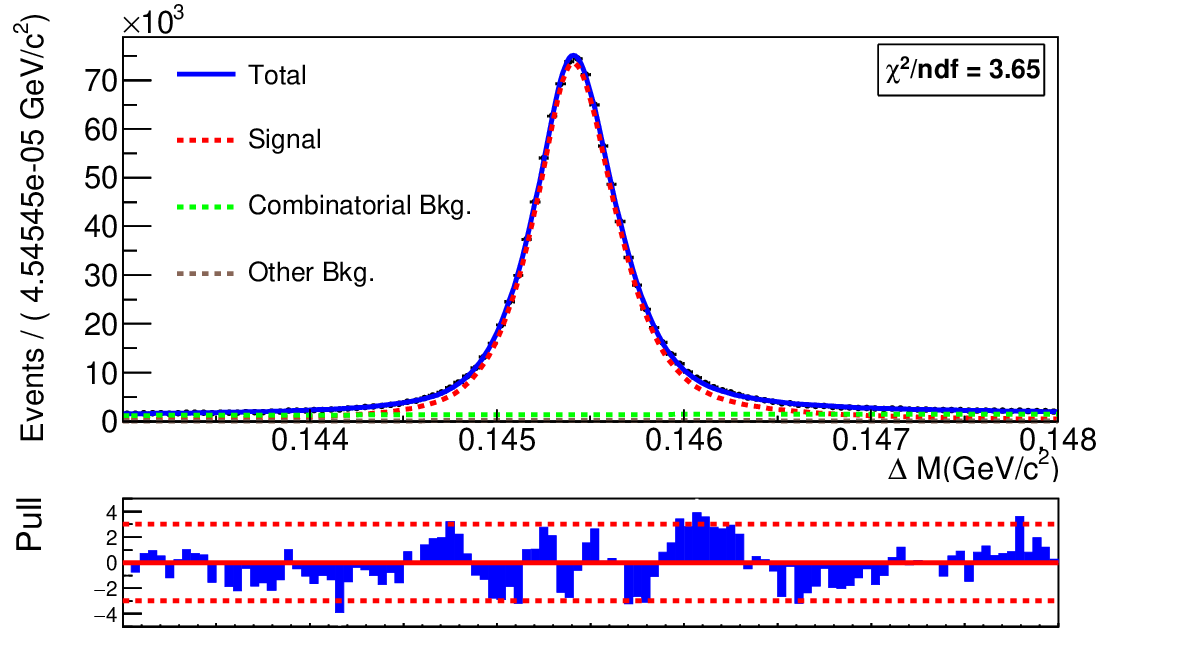}

\caption{Projections of the fit for $\Dkspp$ on $M$ (upper) and $\Delta M$ (lower).
The corresponding pull distributions 
[$= ({\rm data}-{\rm fit\ result})/({\rm data\ uncertainty})$] 
are shown below each projection. The dashed red lines correspond to $\pm 3\sigma$ values.
}
\label{fig:BF_norm}
\end{figure}

We determine $N^{}_{K^0_S\pi^+\pi^-}$ from a two-dimensional
binned fit (rather than unbinned, as the sample is large) 
to the $M$ and $\Delta M$ distributions. The fitted ranges are 
$1.820~{\rm GeV}/c^2 < M < 1.910~{\rm GeV}/c^2$ and 
$0.143~{\rm GeV}/c^2 < \Delta M < 0.148~{\rm GeV}/c^2$~\cite{comment_range}.
We use separate PDFs for correctly reconstructed signal,
slow pion background, broken charm background, and
combinatorial background. 
The small fraction of misreconstructed signal events are 
included in the PDF for combinatorial background.
The functional forms of the
PDFs are mostly the same as those used when fitting 
$\Dkskspp$ events. For the $\Delta M$ PDF for signal, 
the sum of a symmetric Gaussian and an asymmetric Student's~t 
function is used. In addition, the parameter $\sigma^{}_t$
of the Student's~t function is taken to be a function of $M$, 
to account for correlations:
$\sigma^{}_t = \sigma^{}_0 + \sigma^{}_1 (M-m^{}_{D^0})$, 
where $\sigma^{}_0$ and $\sigma^{}_1$ are floated parameters 
and $m^{}_{D^0}$ is the $D^0$ mass~\cite{ParticleDataGroup:2022pth}.
For the $M$ PDF of broken charm background, the sum of 
a Gaussian and a second-order Chebychev polynomial is used. 
For the $M$ PDF of combinatorial 
background, a first-order Chebychev polynomial is used.
There are a total of 10 floated parameters.
The fit yields $1\,069\,870\pm 1831$ $\Dkspp$ decays. 
Projections of the fit are shown in Fig.~\ref{fig:BF_norm}.
The fit quality is somewhat worse than that
for the signal mode due to the very high statistics. We account
for uncertainty in the signal shape when evaluating systematic 
uncertainties (below).

We evaluate the reconstruction efficiencies in 
Eq.~(\ref{eqn:br}) using MC simulation. For 
$\Dkskspp$ decays, no decay model has been measured.
Thus we generate this final state in several ways: 
via four-body phase space, 
via $D^0\ra K^{*+}K^{*-}$ decays,
via $D^0\ra K_S^0\,K_S^0\,\rho^0$ decays,
via $D^0\ra f^0\rho^0$ decays, and
via $D^0\ra K^{*+} K_S^0\,\pi^-$ decays.
The resulting reconstruction efficiencies are found to
span a narrow range; the central value is taken
%in Eq.~(\ref{eqn:br}), 
as our nominal value, and the spread is 
taken as a systematic uncertainty.
%are generated according to four-body phase space, as, at present, 
%no measured decay model is available. 
The $\Dkspp$ decays are generated according to 
the measured Dalitz model~\cite{Belle:2014ydf}. This model includes
$\rho^0\,\overline{K}^{\,0}$,
$\omega \overline{K}^{\,0}$,
$f_0(980) \overline{K}^{\,0}$,
$f_0(1430) \overline{K}^{\,0}$,
$K^*(892)^-\pi^+$,
$K^*_0(1430)^-\pi^+$, and
$K^*_2(1430)^-\pi^+$ intermediate states.
The resulting efficiencies are
$\varepsilon^{}_{K^0_S\,K^0_S\pi^+\pi^-} = (6.92\pm 0.02)\%$ 
and $\varepsilon^{}_{K^0_S\pi^+\pi^-} = (14.88\pm 0.03)\%$,
where the errors are statistical only.
These values are subsequently corrected for small differences 
between data and MC simulation in particle identification (PID) 
and $K^0_S$ reconstruction efficiencies.
The differences are 
measured using $\DstarDpi,\,\Dkp$ and $\DstarDpi,\,\Dkpo$ 
decays, respectively. The overall correction factors are
$0.930\pm 0.014$ for $\Dkskspp$ and 
$0.899\pm 0.007$ for $\Dkspp$.
Inserting all values into Eq.~(\ref{eqn:br}) along with the 
fitted yields and the PDG values~\cite{ParticleDataGroup:2022pth} 
$\mathcal{B}(\Dkspp) = (2.80\pm 0.18)\%$ and
$\mathcal{B}(\KSpp) = (69.20\pm 0.05)\%$ gives 
$\mathcal{B}(\Dkskspp) = (4.79\pm 0.08)\times 10^{-4}$, 
where the quoted uncertainty is statistical only.

The systematic uncertainties on the branching fraction are listed 
in Table~\ref{tab:BF_syst}. The uncertainty arising from the fixed 
parameters in signal and background PDFs is evaluated by varying 
these parameters and refitting. All 31 fixed parameters are sampled 
simultaneously from Gaussian distributions having mean values equal 
to the parameters' nominal values and widths equal to 
their respective uncertainties. After sampling the parameters, the data 
are refit and the resulting signal yield recorded. The procedure is repeated 
5000 times, and the root-mean-square (r.m.s.)\ of the 5000 signal yields 
is taken as the uncertainty due to the fixed parameters. When sampling 
the parameters, correlations among them are accounted for.

The uncertainty due to the fixed yield of broken charm background is 
evaluated by varying this yield (obtained from MC simulation) by $\pm 50$\% 
and refitting. The fractional change in the signal yield is taken as the 
uncertainty. The uncertainty due to the fixed yield of $\Dksksks$ events 
is evaluated in a similar manner; in this case the $\Dksksks$ yield is 
varied by the fractional uncertainty in the branching 
fraction~\cite{ParticleDataGroup:2022pth}. There is a small uncertainty 
due to the finite MC statistics used to evaluate the efficiencies 
$\varepsilon^{}_{K^0_S\,K^0_S\pi^+\pi^-}$ and $\varepsilon^{}_{K^0_S\pi^+\pi^-}$. 

Uncertainty in track reconstruction gives rise to a possible difference 
in reconstruction efficiencies between data and MC simulation. This is evaluated
in a separate study of  $\DstarDpi,\,\Dkspp$ decays~\cite{Dash:2017heu}. 
The resulting uncertainty is 0.35\% per track. As signal decays have two 
more charged tracks than normalization decays do, we take this uncertainty 
to be~0.70\% on the branching fraction. 

There is uncertainty due to $K^0_S$ reconstruction, which is found 
from a study of $\DstarDpi,\,\Dkpo$ decays~\cite{Dash:2017heu}. 
This uncertainty is 0.83\% for $\Dkskspp$ and 0.36\% for $\Dkspp$. 
These uncertainties are correlated between the two channels and
thus partially cancel. However, 
%as the respective $K^0_S$ daughters have different momentum spectra, 
for simplicity we take these uncertainties to be uncorrelated, which is conservative.
The uncertainty due to PID criteria applied to the $\pi^\pm$ 
tracks depends on momentum and is obtained from a study of 
$\DstarDpi,\,\Dkp$ decays. This uncertainty is also correlated 
between the $\Dkskspp$ and $\Dkspp$ channels, and we take this 
correlation into account when calculating the uncertainty.

There is uncertainty arising from the $\Dkspp$ decay model~\cite{Belle:2014ydf}.
We evaluate this uncertainty by modifying the branching 
fractions of intermediate states to correspond to recent 
PDG values~\cite{ParticleDataGroup:2022pth}. These shifts in
intermediate branching fractions are consistent with their statistical
uncertainties. The resulting reconstruction efficiency is 
slightly lower than that of our original decay model; we take 
the average of the two values as our nominal efficiency 
and half the difference as a systematic uncertainty.

There is an uncertainty arising from the $|M(\pi^+\pi^-)-m^{}_{K^0_S}|>10$~MeV/$c^2$ 
requirement applied to reject $\Dksksks$ background. 
This is evaluated by varying this
criterion from 8~MeV/$c^2$ to 15~MeV/$c^2$;
the resulting fractional change in the signal yield is taken as the uncertainty. 
Finally, there is uncertainty in the PDG value $\mathcal{B}(\KSpp)=0.6920\pm 0005$ (which 
enters $\varepsilon$), and the PDG value of the branching fraction for the 
normalization channel $\Dkspp$. The total systematic uncertainty is taken 
as the sum in quadrature of all individual uncertainties. 
The result is $(^{+1.77}_{-1.95})\%$ for $\Dkskspp$, 
$\pm\,0.72\%$ for $\Dkspp$, and $(^{+1.91}_{-2.08})\%$ 
for the ratio of branching fractions.

\begin{table}
%[htb!]
\renewcommand{\arraystretch}{1.3}
\caption{Systematic uncertainties (fractional) for the branching fraction measurement.}
\label{tab:BF_syst}
%\resizebox{\columnwidth}{!}{%
\begin{tabular}{lccc} 
\hline\hline
Source   & $K^0_S\,K^0_S\pi^+\pi^-$  & \hspace*{0.04in} & $K^0_S\pi^+\pi^-$  \\
         &         (\%)             &                  &      (\%)          \\  
\hline
Fixed PDF parameters               &  0.14  &  & 0.09  \\
$\Dksksks$ background              &  0.11  &  &  N/A   \\
Broken charm background            &  0.98  &  &    \\
MC statistics                      &  0.26  &  & 0.17  \\ 
$K^0_S$ reconstruction efficiency  &  0.83  &  & 0.36  \\
PID efficiency correction          & 0.40   &  &       \\
Tracking Efficiency                &  0.70  &  &       \\ 
$M(\pi^{+}\pi^{-})$ veto efficiency &  $^{+0.42}_{-0.93}$ &  & N/A  \\ 
Fraction of misreconstructed signal & $^{+0.02}_{-0.03}$ & &     \\
Decay model                        & 0.73   &  & 0.60   \\ 
$\mathcal{B}(\KSpp)$               & 0.07   &  &        \\ 
\hline
Total for $\mathcal{B}^{}_{K^0_S\,K^0_S\pi^+\pi^-}/\mathcal{B}^{}_{K^0_S\pi^+\pi^-}$  &  
\multicolumn{3}{c}{$^{+1.91}_{-2.08}$}  \\
\hline\hline
\end{tabular}
\end{table}

%%%%%%%%%%%%%%%%%%%%%%%%%%%%%%%%%%%%%%%%%%%%%%%%%%%%%%%%%%%%%%%%

We measure the \cp\ asymmetry $\Acp$ from the difference 
in signal yields for $D^0$ and $\dbar$ decays:
\begin{equation} 
\Acpdet{} = \frac{N(D^{0} \ra f) - N(\dbar\ra \bar{f})}
                 {N(D^{0} \ra f) + N(\dbar\ra \bar{f})}\,.
\label{eqn:acp_def}
\end{equation}
The observable $\Acpdet$ includes asymmetries in production and reconstruction:
\begin{eqnarray}
\Acpdet & = & \Acp  + \Afb + \Aslpi\,,
\end{eqnarray}
where $\Afb$ is the ``forward-backward'' 
production asymmetry~\cite{Afb} between $D^{*+}$ and $D^{*-}$ 
due to $\gamma^{*}-Z^{0}$ interference in $e^{+}e^{-}\ra c\overline{c}$\,; and 
$\Aslpi$ is the asymmetry in reconstruction efficiencies for $\pi^\pm_{s}$ tracks. 
We determine $\Aslpi$ from a study of flavor-tagged $\DstarDpi,\,\Dkp$ decays
and untagged $\Dkp$ decays~\cite{Belle:2011npc}. In this study, 
$\Aslpi$ is measured in bins of $\pt$ and $\costh$ of 
the $\pi^\pm_s$, where $\pt$ is the transverse momentum and
$\theta^{}_{\pi^{}_s}$ is the polar angle 
with respect to the $z$-axis, both evaluated 
in the laboratory frame.
We subsequently correct for $\Aslpi$ in $\kskspp$ events by separately weighting 
$D^0$ and $\dbar$ decays:
\begin{eqnarray}
w^{}_{D^0}   & = & 1 - \Aslpi (\pt,\,\costh) \\
w^{}_{\dbar} & = & 1 + \Aslpi (\pt,\,\costh)\,.
\end{eqnarray}
%where $\Aslpi \Acpwgt$ is the asymmetry in $\pi_{s}$ detection in bins of 
%cosine of $\pi_{s}$ polar angle $cos(\theta_{\pi_{s}})$ and transverse momentum $p_{T}(\pi_{s})$. 

\begin{figure}[ht]
\centering
\includegraphics[width=0.48\textwidth]{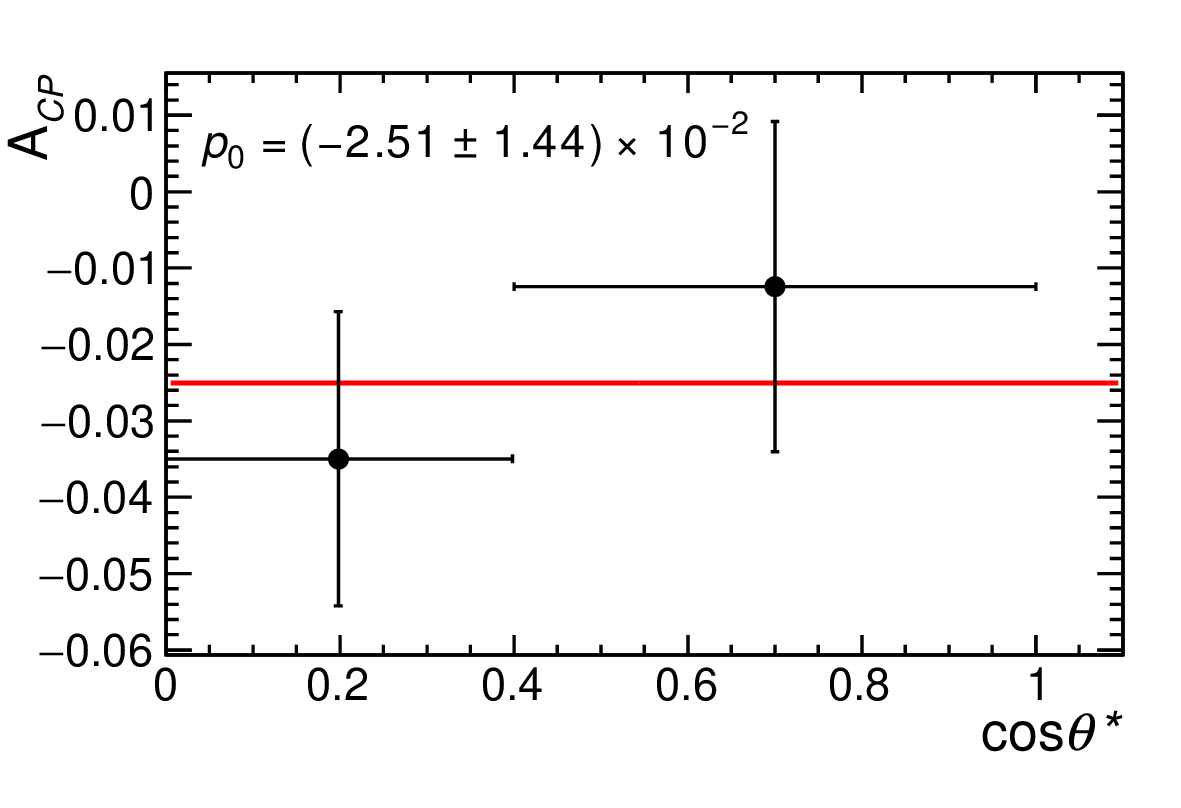}
\vspace{0.5cm}
\includegraphics[width=0.48\textwidth]{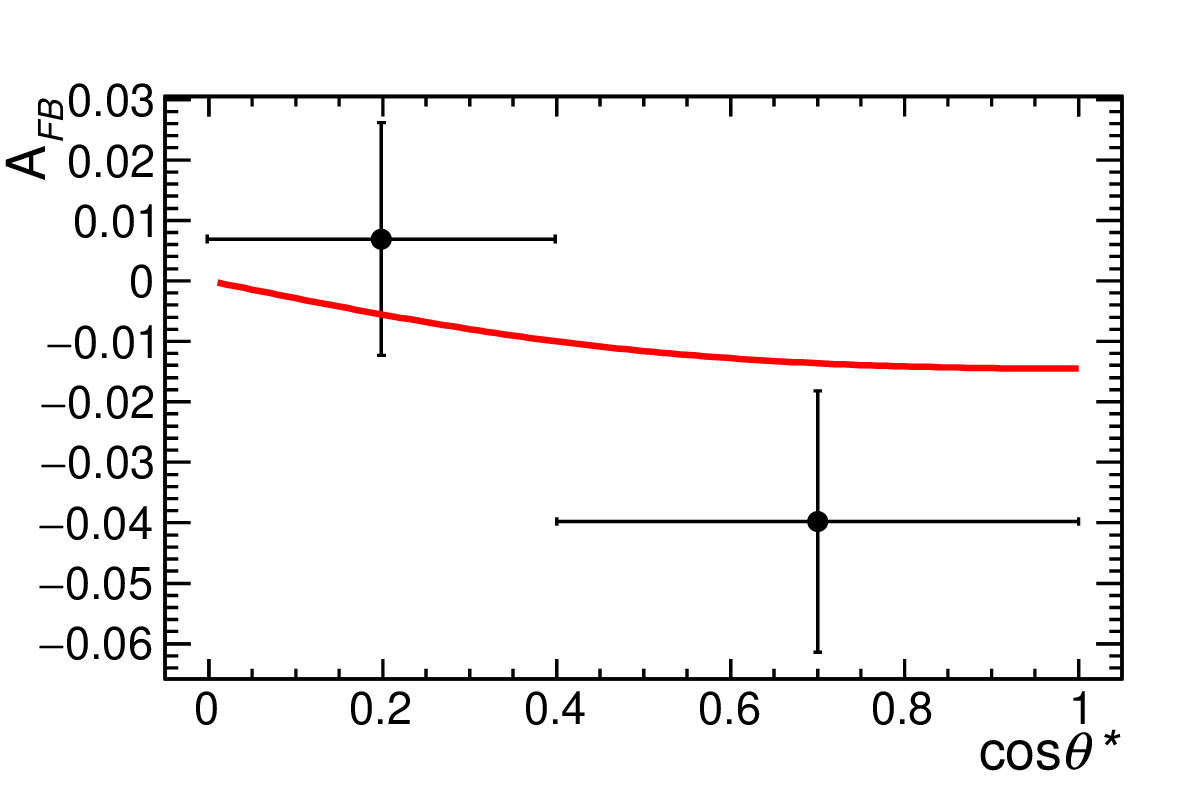}
\caption{Values of $\Acp$ (upper) and $\Afb$ (lower) in
bins of $\cosths$. The red horizontal line in the $\Acp$ plot shows the
result of fitting the points to a constant (``$p_{0}$''). 
The red curve in the $\Afb$ plot shows the leading-order prediction 
for $\Afb (e^{+}e^{-}\rightarrow c\overline{c})$~\cite{pred_afb}.}
\label{fig:Acp_costh}
\end{figure}

After correcting for $\Aslpi$, we obtain $\Acpcor = \Acp + \Afb$. 
The asymmetry $\Afb$ is an odd function of $\cosths$, where $\theta^*$ 
is the polar angle between the $D^{*\pm}$ momentum and the $+z$ axis in 
the CM frame. Since $\Acp$ is a constant, we extract $\Acp$ and also $\Afb$ via
\begin{eqnarray}
\Acp  & =  & \frac{\Acpcor(\cosths) + \Acpcor(-\cosths) }{2} \label{eqn:acp} \\
\Afb  & =  & \frac{\Acpcor(\cosths) - \Acpcor(-\cosths) }{2}\,. \label{eqn:nfb}
\end{eqnarray}
For this calculation, we define four bins of $\cosths$: 
$(-1.0,-0.4)$, $(-0.4, 0)$, $(0, 0.4)$ and $(0.4,1.0)$.
We determine $\Acpcor$ for each bin by simultaneously fitting 
for $D^0$ and $\dbar$ signal yields for weighted events in that bin. 
We use the same PDF functions as used for the branching fraction measurement,
and with the same fixed and floated parameters. The fixed shape parameters 
are taken to be the same for all $\cosths$ bins. The yields of combinatorial 
background for the $D^0$ and $\dbar$ samples are floated independently. 
The yields of broken charm and $\Dksksks$ backgrounds are fixed to MC values. 
The yield of slow pion background is also fixed: the total 
yield is fixed to the value obtained from the branching fraction 
fit, and the fraction assigned to  
$D^0$, $\dbar$, and each $\cosths$ bin is taken from MC simulation.
The fitted parameters are $N(D^0 \ra f)$ and $\Acpcor$. The results 
for $\Acpcor$ are combined according to Eqs.~(\ref{eqn:acp}) and
(\ref{eqn:nfb}) to obtain $\Acp$ and $\Afb$. These values for the
$\cosths$ bins are plotted in Fig.~\ref{fig:Acp_costh}.
Fitting the $\Acp$ values to a constant, we obtain
$\Acp =(-2.51\pm1.44)\%$.

The systematic uncertainties for $\Acp$ are listed in Table~\ref{tab:Acp_syst}.
The uncertainty due to  fixed parameters in the signal and background PDFs is 
evaluated in the same manner as done for the branching fraction: the various
parameters are sampled from Gaussian distributions and the fit is repeated. 
After 2000 trials, the r.m.s.\ of the distribution of $\Acp$ values
is taken as the systematic uncertainty. 

The uncertainty due to the fixed yields of backgrounds is evaluated 
in two ways. The uncertainties in the overall yields of 
broken charm and residual $\Dksksks$ backgrounds are evaluated 
in the same manner as done for the branching fraction measurement. 
In addition, the fixed fractions of the backgrounds between 
$D^0$ and $\dbar$ decays, and among the $\cos\theta^*$ bins, are varied
by sampling these fractions from Gaussian distributions having widths 
equal to the respective uncertainties and repeating the fit. After 
2000 trials, the r.m.s.\ of the resulting distribution of $\Acp$ 
values is again taken as the systematic uncertainty.

We assign a systematic uncertainty due to the choice 
of $\cosths$ binning by generating an ensemble of MC experiments and,
for each experiment, calculating $\Acp$ using four, six, and eight 
bins in $\cosths$. The mean value of $\Acp$ for these bin choices 
is calculated, and the largest difference from the mean value with 
four bins (our nominal result) is taken as the systematic uncertainty.
There is also uncertainty arising 
from the $\Aslpi$ values taken from Ref.~\cite{Belle:2011npc}. We evaluate 
this by sampling $\Aslpi$ values from Gaussian distributions
and refitting for $\Acp$; after 2000 trials, the r.m.s.\ of 
the fitted values is taken as the systematic uncertainty.
The overall systematic uncertainty is the sum in quadrature of 
all individual uncertainties. The result is $(^{+0.11}_{-0.10})\%$.

\begin{table}[bth]
\renewcommand{\arraystretch}{1.4}
\caption{Systematic uncertainties (absolute) for $\Acp$.}
\label{tab:Acp_syst}
\begin{tabular}{lcc} \hline \hline
Sources     & \hspace*{0.20in}  & (\%)           \\  \hline
Fixed PDF parameters      &   &  $\pm 0.01$        \\
$\Dksksks$ background     &   &  $^{+0.02}_{-0.03}$       \\
Broken charm background   &   &  $^{+0.09}_{-0.07}$  \\
Binning in $\cosths$      &   & $\pm 0.04$  \\ 
Reconstruction asymmetry $\Aslpi$ & & $\pm 0.01$  \\ 
Fixed background fractions
%$\Acp$ for peaking backgrounds in $\cosths$ bins} 
& & $\pm 0.04$ \\
\hline
Total                     &   &  $^{+0.11}_{-0.10}$ \\
\hline \hline
\end{tabular}
\end{table}

%%%%%%%%%%%%%%%%%%%%%%%%%%%%%%%%%%%%%%%%%%%%%%%%%%%%%%%%%%%%%%%%

To measure $\acp$, we divide the data into four subsamples:
$D^0$ decays with $C^{}_T>0$ (${\rm yield}\!=\!N^{}_1$) and
$C^{}_T<0$ (${\rm yield}\!=\!N^{}_2$); and 
$\dbar$ decays with $-\overline{C}^{}_T>0$ ($N^{}_3$) and
$-\overline{C}^{}_T<0$ ($N^{}_4$).
Thus, $A^{}_T = (N^{}_1-N^{}_2)/(N^{}_1+N^{}_2)$,
$\bar{A}^{}_T = (N^{}_3-N^{}_4)/(N^{}_3+N^{}_4)$, and 
$\acp = (A^{}_T - \bar{A}^{}_T)/2$.
We fit the four subsamples simultaneously and take the 
fitted parameters to be $N^{}_1$, $N^{}_3$, $A^{}_T$, and~$\acp$. 

\begin{figure*}[h!]
\begin{minipage}[b]{0.35\linewidth}
\includegraphics[width=1.25\textwidth]{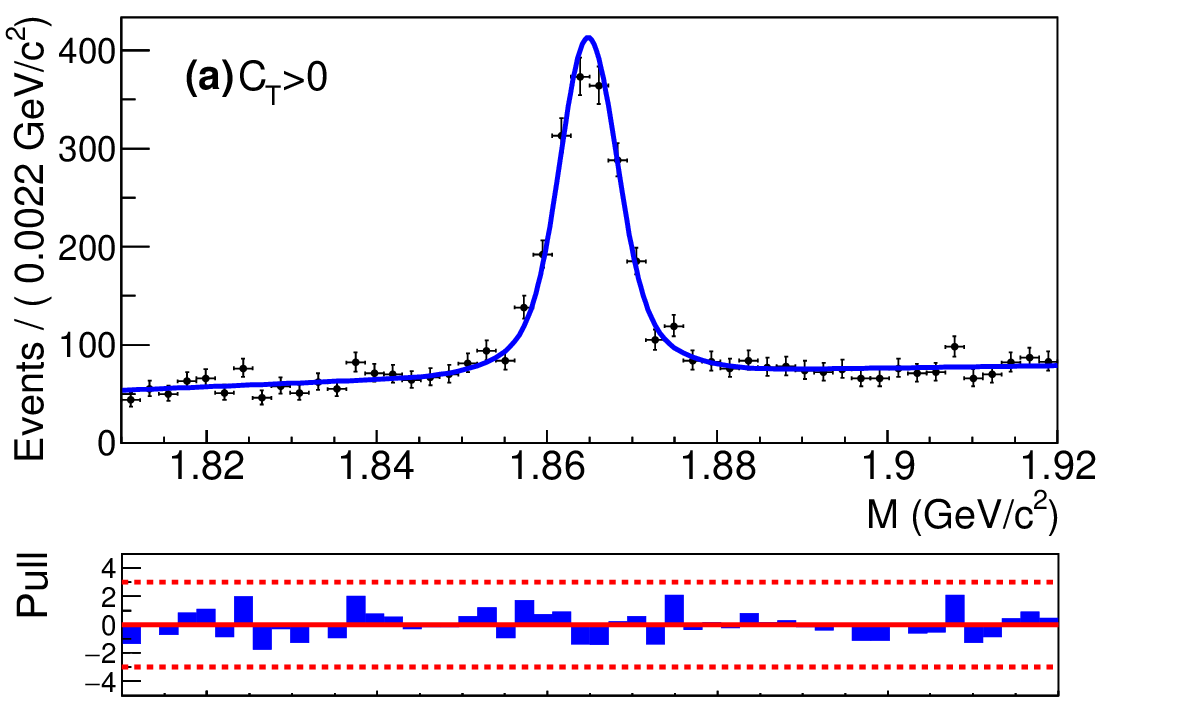}
%\caption{$M_{D^{0}}$ projection}
\end{minipage}
\hspace{0.8cm}
\begin{minipage}[b]{0.35\linewidth}
\includegraphics[width=1.25\textwidth]{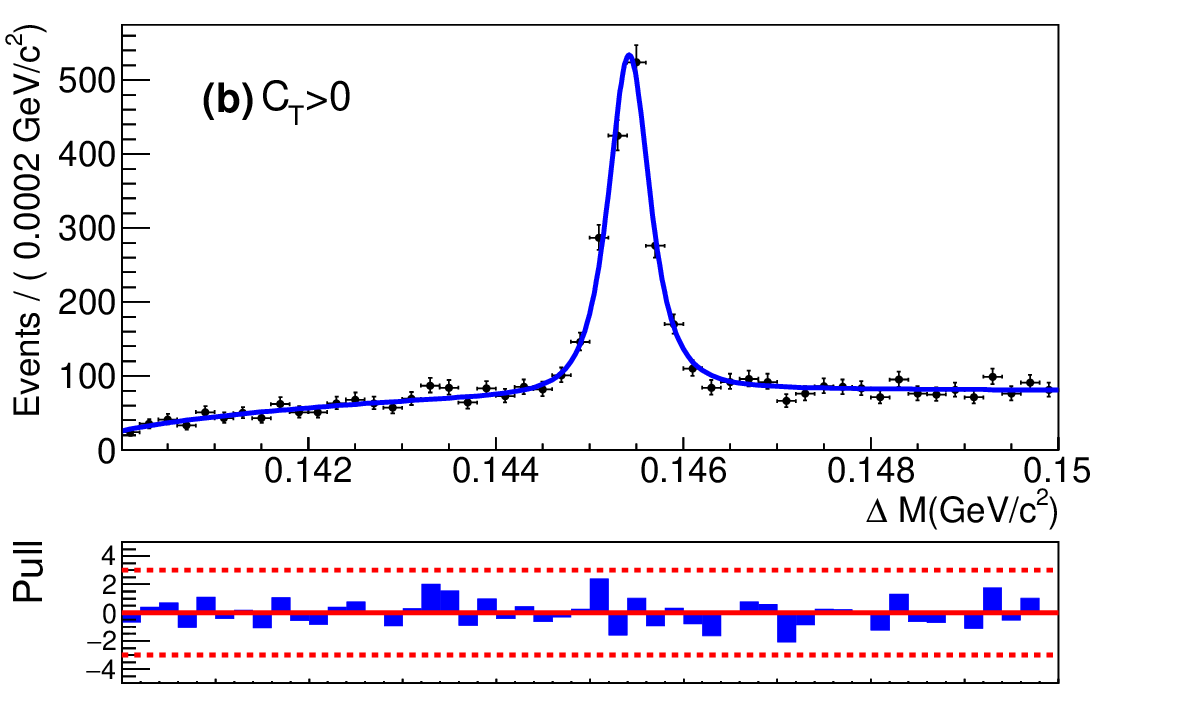}
%\caption{$\Delta M$ projection}
\end{minipage}
\hspace{0.8cm}
\begin{minipage}[b]{0.35\linewidth}
\centering
\includegraphics[width=1.25\textwidth]{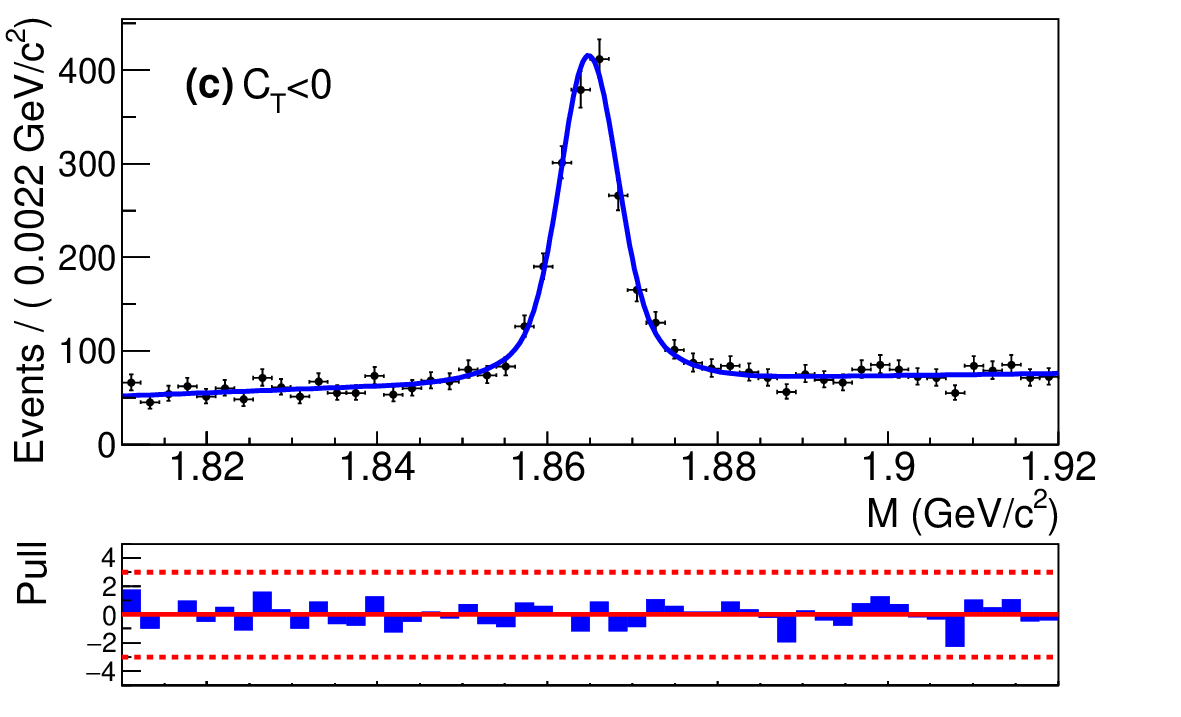}
%\caption{$\Delta M$ projection}
\end{minipage}
\hspace{0.8cm}
\begin{minipage}[b]{0.35\linewidth}
\centering
\includegraphics[width=1.25\textwidth]{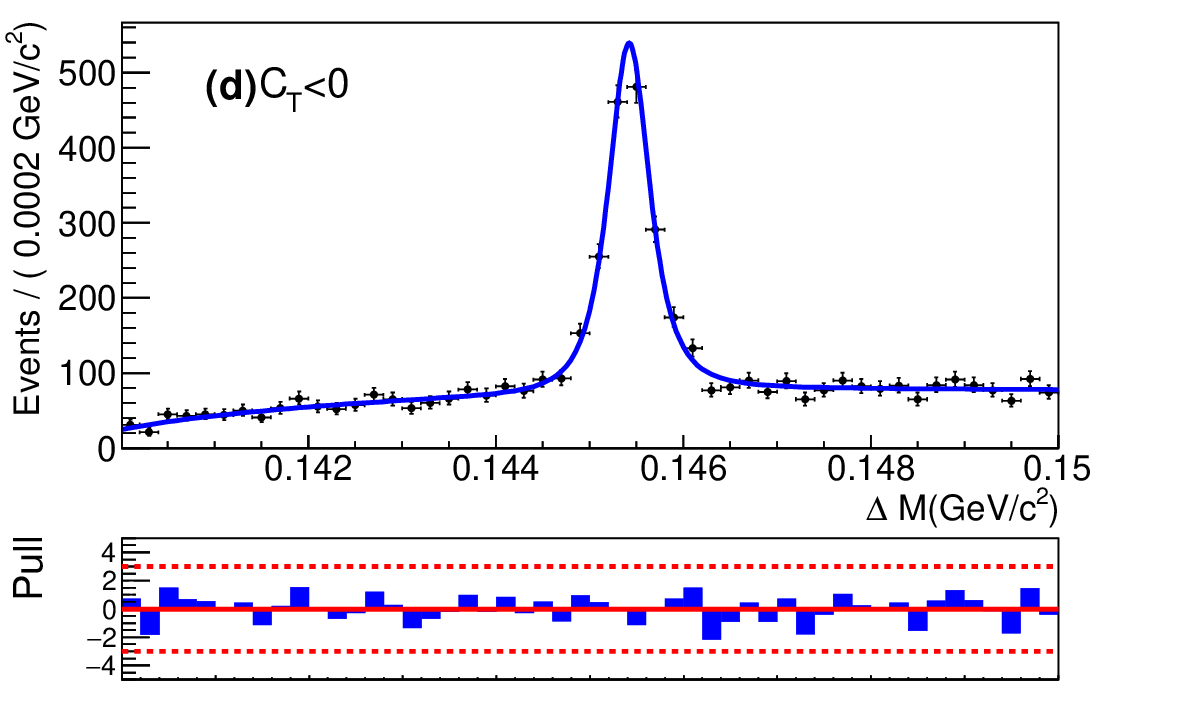}
%\caption{$\Delta M$ projection}
\end{minipage}
\hspace{0.8cm}
\begin{minipage}[b]{0.35\linewidth}
\centering
\includegraphics[width=1.25\textwidth]{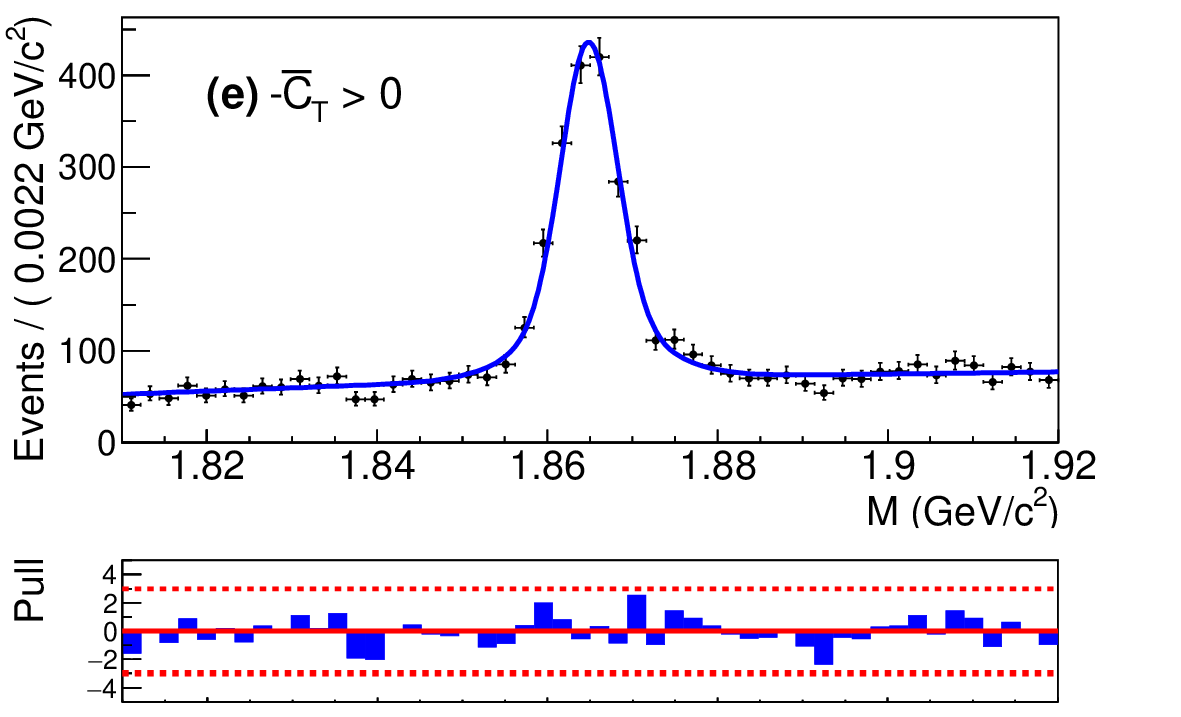}
%\caption{$\Delta M$ projection}
\end{minipage}
\hspace{0.8cm}
\begin{minipage}[b]{0.35\linewidth}
\centering
\includegraphics[width=1.25\textwidth]{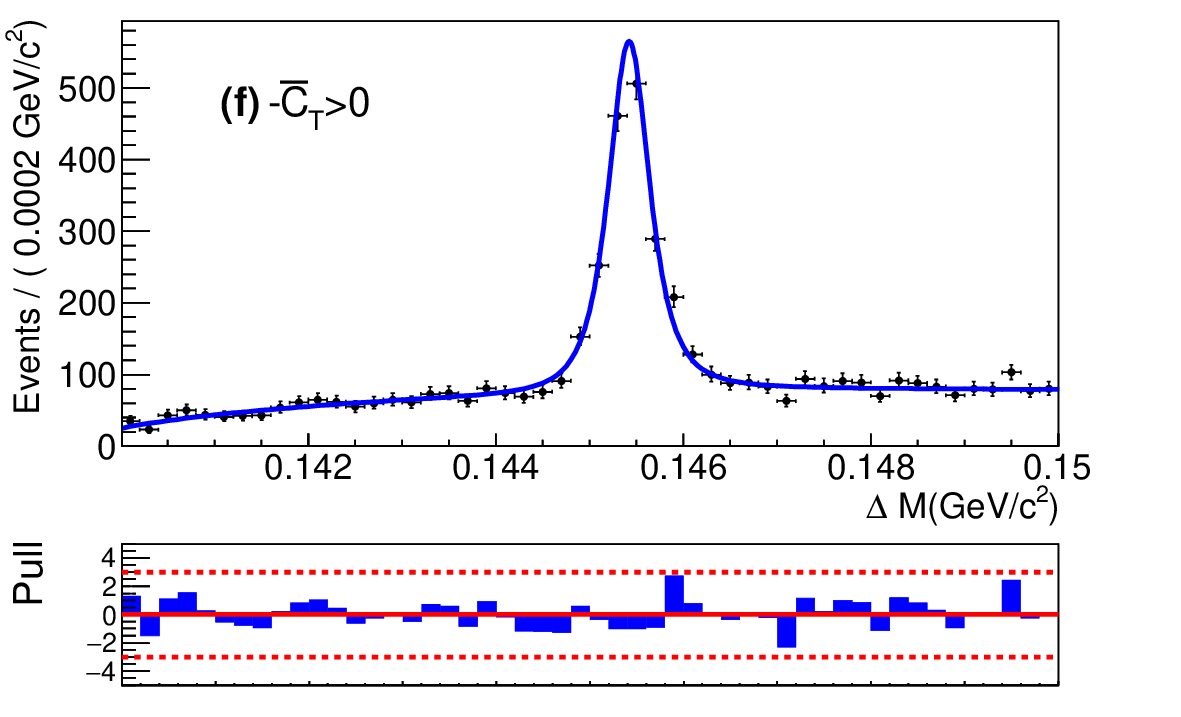}
%\caption{$\Delta M$ projection}
\end{minipage}
\hspace{0.8cm}
\begin{minipage}[b]{0.35\linewidth}
\centering
\includegraphics[width=1.25\textwidth]{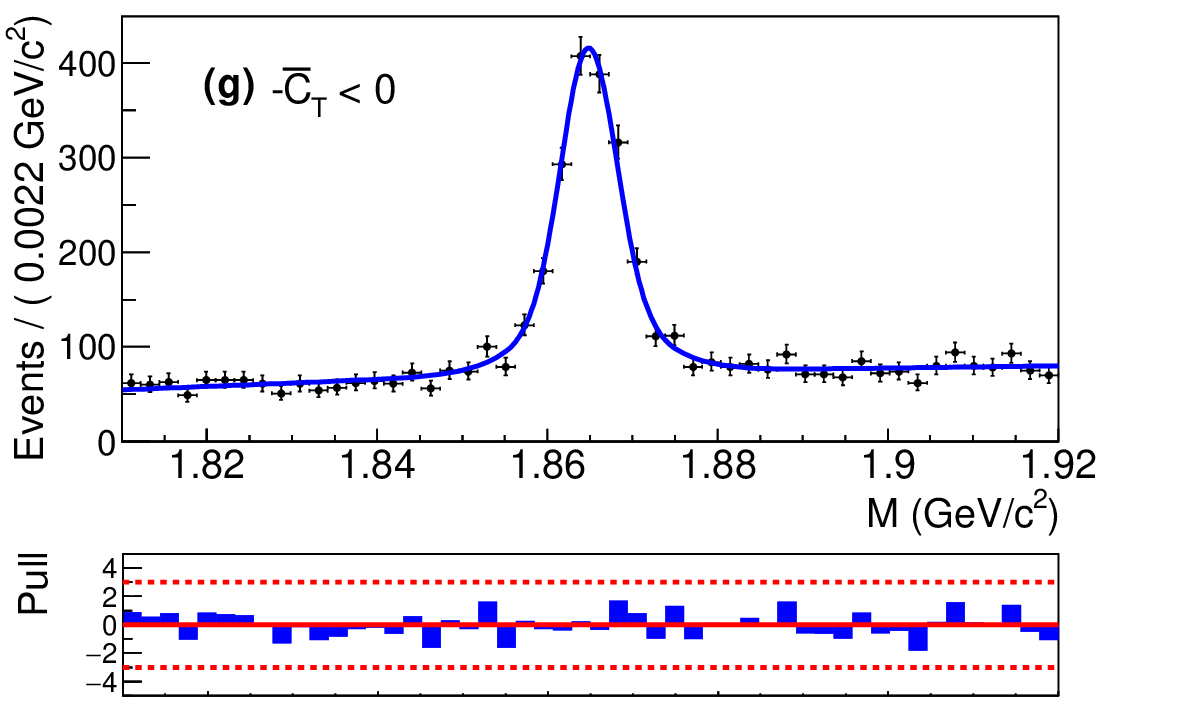}
%\caption{$\Delta M$ projection}
\end{minipage}
\hspace{0.8cm}
\begin{minipage}[b]{0.35\linewidth}
\centering
\includegraphics[width=1.25\textwidth]{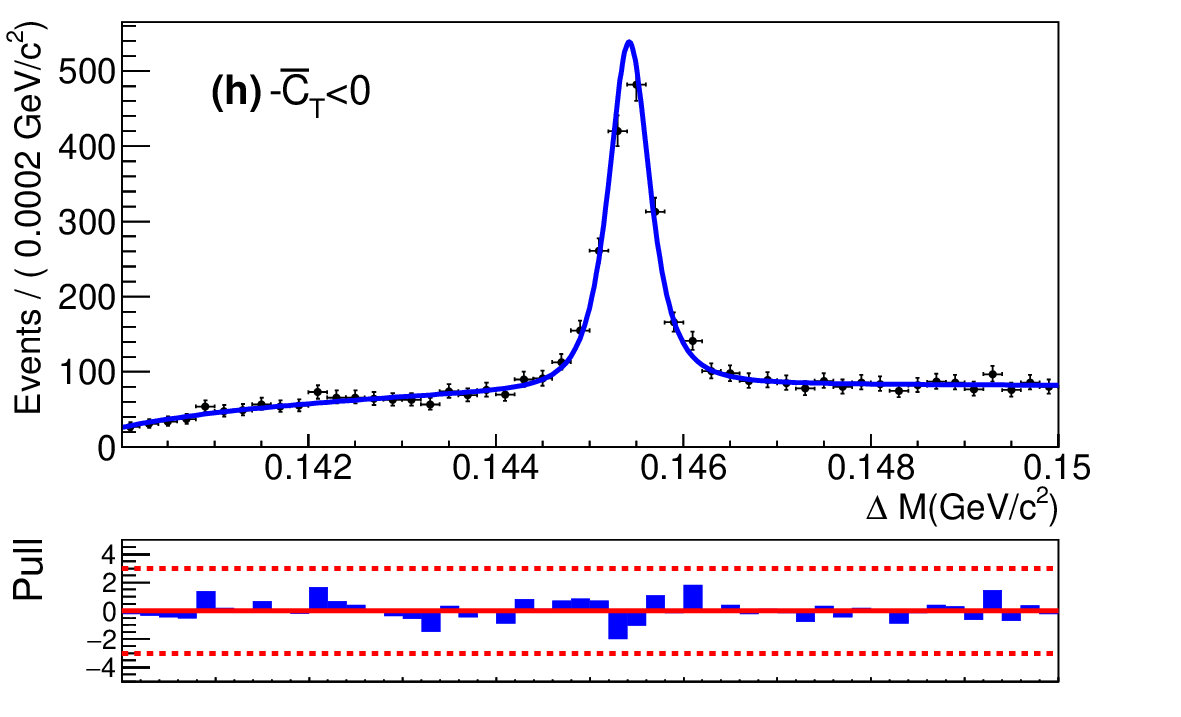}
%\caption{$\Delta M$ projection}
\end{minipage}
\caption{
Projections of the fit for $\acp$ in $M$ (left) and $\Delta M$ (right).
(a) (b) the $D^0$ $C^{}_T>0$ subsample;
(c) (d) the $D^0$ $C^{}_T<0$ subsample;
(e) (f) the $\dbar$ $-\overline{C}^{}_T>0$ subsample; and
(g) (h) the $\dbar$ $-\overline{C}^{}_T<0$ subsample.
}
\label{fig:acp_signal}
\end{figure*}

For this fit, we use the same PDF functions as used for the branching fraction measurement,
and with the same fixed and floated parameters. The fixed shape parameters are taken 
to be the same for all four subsamples, as indicated by MC studies. 
The yield of combinatorial background is floated independently 
for all subsamples. The yield of slow pion background
is fixed in the same way as done for the $\Acp$ fit.
The fit gives 
$A^{}_T = (-0.66\pm 2.01)\%$ and $\acp = (-1.95\pm 1.42)\%$,
where the uncertainties are statistical only.
These values imply $\bar{A}^{}_T = (+3.25\pm 1.98)\%$.
Projections of the fit are shown in Fig.~\ref{fig:acp_signal}.

The systematic uncertainties for $\acp$ are listed in Table~\ref{tab:acp_syst}.
Several uncertainties that enter the branching fraction measurement cancel out for $\acp$.
The uncertainty arising from the fixed parameters in the signal and background PDFs is 
evaluated in the same manner as done for the branching fraction: the various
parameters are sampled from Gaussian distributions, and the fit is repeated. After 
5000 trials, the r.m.s.\ in the fitted values of $\acp$ is taken as the systematic uncertainty. 
The uncertainties due to the fixed yields of broken charm and $\Dksksks$ backgrounds 
are also evaluated in the same manner as done for the branching fraction. 
Finally, we assign an uncertainty due to a possible difference in reconstruction efficiencies between decays with 
$C^{}_T,\,-\overline{C}^{}_T>0$ and those with 
$C^{}_T,\,-\overline{C}^{}_T<0$.
These uncertainties are evaluated using MC simulation
by taking the difference between generated and reconstructed
values of $\acp$.
The total systematic uncertainty is calculated as the sum in quadrature 
of all individual uncertainties; the result is $(^{+0.14}_{-0.12})\%$,
dominated by the uncertainty due to efficiency variation.

\begin{table}
\renewcommand{\arraystretch}{1.4}
\caption{Systematic uncertainties (absolute) for the $\acp$ measurement.}
\label{tab:acp_syst}
\begin{tabular}{lcc} 
\hline\hline
Source                              & \hspace*{0.20in}  &   (\%)  \\  
\hline
Fixed PDF parameters                &    &   0.010                \\
$\Dksksks$ background               &    &  $^{+0.000}_{-0.013}$    \\ 
Broken charm background             &    &  $^{+0.014}_{-0.040}$    \\ 
Efficiency variation 
  with $C^{}_T,\,\overline{C}^{}_T$  &    &  $^{+0.14}_{-0.11}$    \\ 
\hline
Total                               &    &  $^{+0.14}_{-0.12}$    \\
\hline\hline
\end{tabular}
\end{table}

In summary, using Belle data corresponding to an integrated luminosity of 922~fb$^{-1}$,
we measure the branching fraction, $\Acp$, and $\acp$ for $\Dkskspp$ decays. 
The branching fraction, measured relative to that for $\Dkspp$, is
\begin{eqnarray}
\frac{\mathcal{B}(\Dkskspp)}{\mathcal{B}(\Dkspp)} & = & \nonumber \\
 & & \hskip-0.70in 
[1.71\pm 0.03\,({\rm stat})\pm 0.04\,({\rm syst})\,]\times 10^{-2}\,.
\nonumber \\
\end{eqnarray}
Inserting the world average value 
$\mathcal{B}(\Dkspp) = (2.80\pm 0.18)\%$~\cite{ParticleDataGroup:2022pth} gives
\begin{eqnarray}
\mathcal{B}(\Dkskspp) & = & \nonumber \\
 & & 
\hskip-1.40in [4.79\pm 0.08\,({\rm stat})\pm \pm 0.10\,({\rm syst})\pm 0.31\,({\rm norm})]\times 10^{-4}\,,
\nonumber \\
\end{eqnarray}
where the last uncertainty is due to $\mathcal{B}(\Dkspp)$.
The time-integrated \cp\ asymmetry is measured to be
\begin{eqnarray}
\Acp(\Dkskspp) & = & \nonumber \\
 & & \hskip-1.10in 
[-2.51\pm 1.44\,({\rm stat})\,^{+0.11}_{-0.10}\,({\rm syst})]\%\,.
\end{eqnarray}
The \cp-violating asymmetry $\acp$ is measured to be
\begin{eqnarray}
\acp(\Dkskspp) & = & \nonumber \\
 & & \hskip-0.90in [-1.95\,\pm 1.42\,({\rm stat})\,^{+0.14}_{-0.12}\,({\rm syst})] \%\,.
\end{eqnarray}
The branching fraction measurement is the most precise to date. The
measurements of $\Acp$ and $\acp$ are the first such measurements.
We find no evidence of \cp~violation.

\section*{Acknowledgments}
This work, based on data collected using the Belle detector, which was
operated until June 2010, was supported by 
the Ministry of Education, Culture, Sports, Science, and
Technology (MEXT) of Japan, the Japan Society for the 
Promotion of Science (JSPS), and the Tau-Lepton Physics 
Research Center of Nagoya University; 
the Australian Research Council including grants
No.~DP180102629, % Sevior
No.~DP170102389, % Varvell
No.~DP170102204, % Yabsley
No.~DE220100462, % Hsu
No.~DP150103061, % Urquijo
No.~FT130100303; % Urquijo;
Austrian Federal Ministry of Education, Science and Research (FWF) and
FWF Austrian Science Fund No.~P~31361-N36;
the National Natural Science Foundation of China under Contracts
No.~11675166,  %Wen-Biao Yan
No.~11705209;  %Yi-Ming Li
No.~11975076;  %Chengping Shen
No.~12135005;  %Chengping Shen 
No.~12175041;  %Xiaolong Wang
No.~12161141008; %Chengping Shen
Key Research Program of Frontier Sciences, Chinese Academy of Sciences (CAS), Grant No.~QYZDJ-SSW-SLH011; % Chang-Zheng Yuan
Project ZR2022JQ02 supported by Shandong Provincial Natural Science Foundation;
the Ministry of Education, Youth and Sports of the Czech
Republic under Contract No.~LTT17020;
the Czech Science Foundation Grant No. 22-18469S;
Horizon 2020 ERC Advanced Grant No.~884719 and ERC Starting Grant No.~947006 ``InterLeptons'' (European Union);
the Carl Zeiss Foundation, the Deutsche Forschungsgemeinschaft, the
Excellence Cluster Universe, and the VolkswagenStiftung;
the Department of Atomic Energy (Project Identification No. RTI 4002) and the Department of Science and Technology of India; 
the Istituto Nazionale di Fisica Nucleare of Italy; 
National Research Foundation (NRF) of Korea 
Grants No.~2016R1\-D1A1B\-02012900, 
No.~2018R1\-A2B\-3003643,
No.~2018R1\-A6A1A\-06024970, 
No.~RS\-2022\-00197659,
No.~2019R1\-I1A3A\-01058933, 
No.~2021R1\-A6A1A\-03043957,
No.~2021R1\-F1A\-1060423, 
No.~2021R1\-F1A\-1064008, 
No.~2022R1\-A2C\-1003993;
Radiation Science Research Institute, Foreign Large-size Research Facility Application Supporting project, 
the Global Science Experimental Data Hub Center of the Korea Institute of Science and Technology Information 
and KREONET/GLORIAD;
the Polish Ministry of Science and Higher Education and 
the National Science Center;
the Ministry of Science and Higher Education of the Russian Federation, Agreement 14.W03.31.0026, % from 15.02.2018
and the HSE University Basic Research Program, Moscow; % from 15.04.2021
University of Tabuk research grants
No.~S-1440-0321, 
No.~S-0256-1438, and 
No.~S-0280-1439 (Saudi Arabia);
the Slovenian Research Agency Grants 
No.~J1-9124 and No.~P1-0135;
Ikerbasque, Basque Foundation for Science, Spain;
the Swiss National Science Foundation; 
the Ministry of Education and the Ministry of Science and Technology of Taiwan;
and the United States Department of Energy and the National Science Foundation.
These acknowledgements are not to be interpreted as an endorsement of any
statement made by any of our institutes, funding agencies, governments, or
their representatives.
We thank the KEKB group for the excellent operation of the
accelerator; the KEK cryogenics group for the efficient
operation of the solenoid; and the KEK computer group and the Pacific Northwest National
Laboratory (PNNL) Environmental Molecular Sciences Laboratory (EMSL)
computing group for strong computing support; and the National
Institute of Informatics, and Science Information NETwork 6 (SINET6) for
valuable network support.

\bibliographystyle{apsrev}
\bibliography{references}

\begin{thebibliography}{34}
\expandafter\ifx\csname natexlab\endcsname\relax\def\natexlab#1{#1}\fi
\expandafter\ifx\csname bibnamefont\endcsname\relax
  \def\bibnamefont#1{#1}\fi
\expandafter\ifx\csname bibfnamefont\endcsname\relax
  \def\bibfnamefont#1{#1}\fi
\expandafter\ifx\csname citenamefont\endcsname\relax
  \def\citenamefont#1{#1}\fi
\expandafter\ifx\csname url\endcsname\relax
  \def\url#1{\texttt{#1}}\fi
\expandafter\ifx\csname urlprefix\endcsname\relax\def\urlprefix{URL }\fi
\providecommand{\bibinfo}[2]{#2}
\providecommand{\eprint}[2][]{\url{#2}}

\bibitem[{\citenamefont{Canetti et~al.}(2012)\citenamefont{Canetti, Drewes, and
  Shaposhnikov}}]{Canetti:2012zc}
\bibinfo{author}{\bibfnamefont{L.}~\bibnamefont{Canetti}},
  \bibinfo{author}{\bibfnamefont{M.}~\bibnamefont{Drewes}}, \bibnamefont{and}
  \bibinfo{author}{\bibfnamefont{M.}~\bibnamefont{Shaposhnikov}},
  \bibinfo{journal}{New J. Phys.} \textbf{\bibinfo{volume}{14}},
  \bibinfo{pages}{095012} (\bibinfo{year}{2012}).

\bibitem[{\citenamefont{Farrar and Shaposhnikov}(1994)}]{Farrar:1993hn}
\bibinfo{author}{\bibfnamefont{G.~R.} \bibnamefont{Farrar}} \bibnamefont{and}
  \bibinfo{author}{\bibfnamefont{M.~E.} \bibnamefont{Shaposhnikov}},
  \bibinfo{journal}{Phys. Rev. D} \textbf{\bibinfo{volume}{50}},
  \bibinfo{pages}{774} (\bibinfo{year}{1994}).

\bibitem[{\citenamefont{Allahverdi et~al.}(2021)}]{Allahverdi_2021}
\bibinfo{author}{\bibfnamefont{R.}~\bibnamefont{Allahverdi}}
  \bibnamefont{et~al.}, \bibinfo{journal}{Open J. Astrophys.}
  \textbf{\bibinfo{volume}{4}} (\bibinfo{year}{2021}).

\bibitem[{\citenamefont{Sakharov}(1991)}]{Sakharov:1967dj}
\bibinfo{author}{\bibfnamefont{A.~D.} \bibnamefont{Sakharov}},
  \bibinfo{journal}{Sov. Phys. Usp.} \textbf{\bibinfo{volume}{34}},
  \bibinfo{pages}{392} (\bibinfo{year}{1991}).

\bibitem[{\citenamefont{Kobayashi and Maskawa}(1973)}]{Kobayashi:1973fv}
\bibinfo{author}{\bibfnamefont{M.}~\bibnamefont{Kobayashi}} \bibnamefont{and}
  \bibinfo{author}{\bibfnamefont{T.}~\bibnamefont{Maskawa}},
  \bibinfo{journal}{Prog. Theor. Phys.} \textbf{\bibinfo{volume}{49}},
  \bibinfo{pages}{652} (\bibinfo{year}{1973}).

\bibitem[{\citenamefont{Huet and Sather}(1995)}]{Huet:1994jb}
\bibinfo{author}{\bibfnamefont{P.}~\bibnamefont{Huet}} \bibnamefont{and}
  \bibinfo{author}{\bibfnamefont{E.}~\bibnamefont{Sather}},
  \bibinfo{journal}{Phys. Rev. D} \textbf{\bibinfo{volume}{51}},
  \bibinfo{pages}{379} (\bibinfo{year}{1995}).

\bibitem[{cha()}]{charge-conjugates}
\bibinfo{note}{Charge-conjugate modes are implicitly included unless noted
  otherwise.}

\bibitem[{\citenamefont{Grossman et~al.}(2007)\citenamefont{Grossman, Kagan,
  and Nir}}]{Grossman:2006jg}
\bibinfo{author}{\bibfnamefont{Y.}~\bibnamefont{Grossman}},
  \bibinfo{author}{\bibfnamefont{A.~L.} \bibnamefont{Kagan}}, \bibnamefont{and}
  \bibinfo{author}{\bibfnamefont{Y.}~\bibnamefont{Nir}},
  \bibinfo{journal}{Phys. Rev. D} \textbf{\bibinfo{volume}{75}},
  \bibinfo{pages}{036008} (\bibinfo{year}{2007}).

\bibitem[{\citenamefont{Aaij et~al.}(2019)}]{LHCb:2019hro}
\bibinfo{author}{\bibfnamefont{R.}~\bibnamefont{Aaij}} \bibnamefont{et~al.}
  (\bibinfo{collaboration}{LHCb Collaboration}), \bibinfo{journal}{Phys. Rev.
  Lett.} \textbf{\bibinfo{volume}{122}}, \bibinfo{pages}{211803}
  (\bibinfo{year}{2019}).

\bibitem[{\citenamefont{Albrecht et~al.}(1994)}]{Albrecht:1994}
\bibinfo{author}{\bibfnamefont{H.}~\bibnamefont{Albrecht}} \bibnamefont{et~al.}
  (\bibinfo{collaboration}{ARGUS Collaboration}), \bibinfo{journal}{Zeit. Phys.
  C} \textbf{\bibinfo{volume}{64}}, \bibinfo{pages}{375}
  (\bibinfo{year}{1994}).

\bibitem[{\citenamefont{Link et~al.}(2005)}]{FOCUS:2004met}
\bibinfo{author}{\bibfnamefont{J.~M.} \bibnamefont{Link}} \bibnamefont{et~al.}
  (\bibinfo{collaboration}{FOCUS Collaboration}), \bibinfo{journal}{Phys. Lett.
  B} \textbf{\bibinfo{volume}{607}}, \bibinfo{pages}{59}
  (\bibinfo{year}{2005}).

\bibitem[{\citenamefont{Ablikim et~al.}(2020)}]{BESIII:2020rxv}
\bibinfo{author}{\bibfnamefont{M.}~\bibnamefont{Ablikim}} \bibnamefont{et~al.}
  (\bibinfo{collaboration}{BESIII Collaboration}), \bibinfo{journal}{Phys. Rev.
  D} \textbf{\bibinfo{volume}{102}}, \bibinfo{pages}{052006}
  (\bibinfo{year}{2020}).

\bibitem[{\citenamefont{Durieux and Grossman}(2015)}]{Durieux:2015zwa}
\bibinfo{author}{\bibfnamefont{G.}~\bibnamefont{Durieux}} \bibnamefont{and}
  \bibinfo{author}{\bibfnamefont{Y.}~\bibnamefont{Grossman}},
  \bibinfo{journal}{Phys. Rev. D} \textbf{\bibinfo{volume}{92}},
  \bibinfo{pages}{076013} (\bibinfo{year}{2015}).

\bibitem[{\citenamefont{Valencia}(1989)}]{PhysRevD.39.3339}
\bibinfo{author}{\bibfnamefont{G.}~\bibnamefont{Valencia}},
  \bibinfo{journal}{Phys. Rev. D} \textbf{\bibinfo{volume}{39}},
  \bibinfo{pages}{3339} (\bibinfo{year}{1989}).

\bibitem[{\citenamefont{Bensalem and London}(2001)}]{Bensalem:2000hq}
\bibinfo{author}{\bibfnamefont{W.}~\bibnamefont{Bensalem}} \bibnamefont{and}
  \bibinfo{author}{\bibfnamefont{D.}~\bibnamefont{London}},
  \bibinfo{journal}{Phys. Rev. D} \textbf{\bibinfo{volume}{64}},
  \bibinfo{pages}{116003} (\bibinfo{year}{2001}).

\bibitem[{\citenamefont{Kurokawa and Kikutani}(2003)}]{kekb}
\bibinfo{author}{\bibfnamefont{S.}~\bibnamefont{Kurokawa}} \bibnamefont{and}
  \bibinfo{author}{\bibfnamefont{E.}~\bibnamefont{Kikutani}},
  \bibinfo{journal}{Nucl. Instrum. Methods Phys. Res., Sect. A}
  \textbf{\bibinfo{volume}{499}}, \bibinfo{pages}{1} (\bibinfo{year}{2003}),
  \bibinfo{note}{and other papers in this volume; T.~Abe {\it et al.}, Prog.
  Theor. Exp. Phys. {\bf 2013}, 03A001 (2013), and references therein.}

\bibitem[{\citenamefont{Abashian et~al.}(2002)}]{Abashian:2000cg}
\bibinfo{author}{\bibfnamefont{A.}~\bibnamefont{Abashian}} \bibnamefont{et~al.}
  (\bibinfo{collaboration}{Belle Collaboration}), \bibinfo{journal}{Nucl.
  Instrum. Methods Phys. Res., Sect. A} \textbf{\bibinfo{volume}{479}},
  \bibinfo{pages}{117} (\bibinfo{year}{2002}), \bibinfo{note}{also see Section
  2 in J.~Brodzicka {\it et al.}, Prog. Theor. Exp. Phys. {\bf 2012}, 04D001
  (2012).}

\bibitem[{\citenamefont{Natkaniec et~al.}(2006)}]{Natkaniec:2006rv}
\bibinfo{author}{\bibfnamefont{Z.}~\bibnamefont{Natkaniec}}
  \bibnamefont{et~al.}, \bibinfo{journal}{Nucl. Instrum. Methods Phys. Res.,
  Sect. A} \textbf{\bibinfo{volume}{560}}, \bibinfo{pages}{1}
  (\bibinfo{year}{2006}).

\bibitem[{\citenamefont{Lange}(2001)}]{Lange:2001uf}
\bibinfo{author}{\bibfnamefont{D.~J.} \bibnamefont{Lange}},
  \bibinfo{journal}{Nucl. Instrum. Methods Phys. Res., Sect. A}
  \textbf{\bibinfo{volume}{462}}, \bibinfo{pages}{152} (\bibinfo{year}{2001}).

\bibitem[{\citenamefont{Brun et~al.}(1987)}]{Brun:1987ma}
\bibinfo{author}{\bibfnamefont{R.}~\bibnamefont{Brun}} \bibnamefont{et~al.},
  \bibinfo{journal}{GEANT 3.21, Report No: CERN Report DD/EE/84-1}
  (\bibinfo{year}{1987}).

\bibitem[{\citenamefont{Barberio and Was}(1994)}]{Barberio:1993qi}
\bibinfo{author}{\bibfnamefont{E.}~\bibnamefont{Barberio}} \bibnamefont{and}
  \bibinfo{author}{\bibfnamefont{Z.}~\bibnamefont{Was}},
  \bibinfo{journal}{Comput. Phys. Commun.} \textbf{\bibinfo{volume}{79}},
  \bibinfo{pages}{291} (\bibinfo{year}{1994}).

\bibitem[{\citenamefont{Dash et~al.}(2017)}]{Dash:2017heu}
\bibinfo{author}{\bibfnamefont{N.}~\bibnamefont{Dash}} \bibnamefont{et~al.}
  (\bibinfo{collaboration}{Belle Collaboration}), \bibinfo{journal}{Phys. Rev.
  Lett.} \textbf{\bibinfo{volume}{119}}, \bibinfo{pages}{171801}
  (\bibinfo{year}{2017}).

\bibitem[{\citenamefont{Workman et~al.}(2022)}]{ParticleDataGroup:2022pth}
\bibinfo{author}{\bibfnamefont{R.~L.} \bibnamefont{Workman}}
  \bibnamefont{et~al.} (\bibinfo{collaboration}{Particle Data Group}),
  \bibinfo{journal}{Prog. Theor. Exp. Phys.} \textbf{\bibinfo{volume}{2022}},
  \bibinfo{pages}{083C01} (\bibinfo{year}{2022}).

\bibitem[{\citenamefont{Aubert et~al.}(2009)}]{BaBar:2008xnt}
\bibinfo{author}{\bibfnamefont{B.}~\bibnamefont{Aubert}} \bibnamefont{et~al.}
  (\bibinfo{collaboration}{{\it BABAR\/} Collaboration}),
  \bibinfo{journal}{Phys. Rev. D} \textbf{\bibinfo{volume}{{\bf{79}}}},
  \bibinfo{pages}{032002} (\bibinfo{year}{2009}).

\bibitem[{\citenamefont{Lees et~al.}(2012)}]{BaBar:2012cyz}
\bibinfo{author}{\bibfnamefont{J.~P.} \bibnamefont{Lees}} \bibnamefont{et~al.}
  (\bibinfo{collaboration}{{\it BABAR\/} Collaboration}),
  \bibinfo{journal}{Phys. Rev. D} \textbf{\bibinfo{volume}{86}},
  \bibinfo{pages}{112006} (\bibinfo{year}{2012}).

\bibitem[{\citenamefont{Kronenbitter et~al.}(2012)}]{Belle:2012xkw}
\bibinfo{author}{\bibfnamefont{B.}~\bibnamefont{Kronenbitter}}
  \bibnamefont{et~al.} (\bibinfo{collaboration}{Belle Collaboration}),
  \bibinfo{journal}{Phys. Rev. D} \textbf{\bibinfo{volume}{86}},
  \bibinfo{pages}{071103} (\bibinfo{year}{2012}).

\bibitem[{\citenamefont{Rohrken et~al.}(2012)}]{Belle:2012mef}
\bibinfo{author}{\bibfnamefont{M.}~\bibnamefont{Rohrken}} \bibnamefont{et~al.}
  (\bibinfo{collaboration}{Belle Collaboration}), \bibinfo{journal}{Phys. Rev.
  D} \textbf{\bibinfo{volume}{85}}, \bibinfo{pages}{091106}
  (\bibinfo{year}{2012}).

\bibitem[{\citenamefont{Aaij et~al.}(2020)}]{LHCb:2019ouq}
\bibinfo{author}{\bibfnamefont{R.}~\bibnamefont{Aaij}} \bibnamefont{et~al.}
  (\bibinfo{collaboration}{LHCb Collaboration}), \bibinfo{journal}{J. High
  Energy Phys.} \textbf{\bibinfo{volume}{03}}, \bibinfo{pages}{147}
  (\bibinfo{year}{2020}).

\bibitem[{\citenamefont{James}(2006)}]{student-t}
\bibinfo{author}{\bibfnamefont{F.~E.} \bibnamefont{James}},
  \emph{\bibinfo{title}{{Statistical Methods in Experimental Physics 2nd ed.}}}
  (\bibinfo{publisher}{World Scientific}, \bibinfo{address}{Singapore},
  \bibinfo{year}{2006}), \bibinfo{note}{pp. 73-75}.

\bibitem[{com()}]{comment_range}
\bibinfo{note}{The fitted ranges are larger for the signal mode in order to
  more accurately model the background level.}

\bibitem[{\citenamefont{Peng et~al.}(2014)}]{Belle:2014ydf}
\bibinfo{author}{\bibfnamefont{T.}~\bibnamefont{Peng}} \bibnamefont{et~al.}
  (\bibinfo{collaboration}{Belle Collaboration}), \bibinfo{journal}{Phys. Rev.
  D} \textbf{\bibinfo{volume}{89}}, \bibinfo{pages}{091103}
  (\bibinfo{year}{2014}), \eprint{1404.2412}.

\bibitem[{\citenamefont{Berends et~al.}(1973)\citenamefont{Berends, Gaemers,
  and Gastmans}}]{Afb}
\bibinfo{author}{\bibfnamefont{F.}~\bibnamefont{Berends}},
  \bibinfo{author}{\bibfnamefont{K.}~\bibnamefont{Gaemers}}, \bibnamefont{and}
  \bibinfo{author}{\bibfnamefont{R.}~\bibnamefont{Gastmans}},
  \bibinfo{journal}{Nucl. Phys.} \textbf{\bibinfo{volume}{B63}},
  \bibinfo{pages}{381} (\bibinfo{year}{1973}), \bibinfo{note}{also see
  R.~J.~Cashmore, C.~M.~Hawkes, B.~W.~Lynn, and R.~G.~Stuart, Z.~Phys.~C {\bf
  30}, 125 (1986)}.

\bibitem[{\citenamefont{Ko et~al.}(2011)}]{Belle:2011npc}
\bibinfo{author}{\bibfnamefont{B.~R.} \bibnamefont{Ko}} \bibnamefont{et~al.}
  (\bibinfo{collaboration}{Belle Collaboration}), \bibinfo{journal}{Phys. Rev.
  Lett.} \textbf{\bibinfo{volume}{106}}, \bibinfo{pages}{211801}
  (\bibinfo{year}{2011}).

\bibitem[{pre()}]{pred_afb}
\bibinfo{note}{The leading-order prediction at $\sqrt{s} = 10.6$~GeV is
  $\Afb(e^+e^- \rightarrow c\bar{c})\approx
  -0.029\cos\theta^*/(1+\cos^2\theta^*)$; see, for example, O.~Nachtmann, {\it
  Elementary Particle Physics\/} (Springer-Verlag, Berlin, 1989).}

\end{thebibliography}
\end{document}